\documentclass[lettersize,journal]{IEEEtran}

\usepackage[utf8]{inputenc}
\usepackage{graphicx}
\usepackage{amsmath, amssymb}
\usepackage{hyperref}
\usepackage{amsfonts} 
\usepackage{bm}
\usepackage{algorithm}
\usepackage{algorithmic}
\usepackage{caption} 
\usepackage{cite}
\newcounter{tempfig}
\begin{document}
\title{Secrecy Energy Efficiency for 

IRS-Assisted Low-Altitude Communications: 

A D3QN-PER Based Approach}
\author{Ya Gao, Peina Zhao, Yiheng Li, Wenchi Cheng
\thanks{}
\thanks{Ya Gao is with School of Physics and Electronic Information, Luoyang Normal University, Luoyang, 471934, China, and also with Xidian University, Xi'an 710071, China (e-mail: gaoya@lynu.edu.cn).}
\thanks{Peina Zhao and Wenchi Cheng are with Xidian University, Xi’an 710071, China (e-mail: peinazhao@stu.xidian.edu.cn; wccheng@xidian.edu.cn).}
\thanks{ Yiheng Li is with School of Physics and Electronic Information, Luoyang Normal University, Luoyang, 471934, China.}
}
\markboth{}
{Gao \MakeLowercase{\textit{et al.}}: A Sample Article Using IEEEtran.cls for IEEE Journals}


\maketitle
\begin{abstract}
To address the security and energy efficiency challenges in low-altitude economy (LAE) wireless communications, we develop a secure synergistic network integrating unmanned aerial vehicle (UAV) and intelligent reflecting surface (IRS), with an emphasis on maximizing secrecy energy efficiency (SEE) for downlink transmission scenarios. In particular, firstly, we establish the channel transmission models for UAV-IRS assisted LAE communications network. 
Then, we formulate a non-convex fractional optimization problem for SEE maximization, involving three tightly coupled variables, i.e. the beamforming, IRS phase and UAV trajectory. To tackle the fractional structure and variable coupling, Dinkelbach’s method and equivalent transformations are leveraged to reformulate the objective function, which is then decoupled and decomposed into three independent subproblems via an alternating optimization strategy for iterative resolution. 
Slack variables and Semidefinite Relaxation (SDR) are further employed to convexify the subproblems of beamforming and IRS phase shift optimization, thereby obtaining their optimal solutions. 
For the UAV trajectory optimization subproblem, we propose a D3QN-PER algorithm, which integrates a Dueling Double Deep Q-Network with Prioritized Experience Replay, to tackle the slow convergence and training instability inherent in conventional Deep Q-Network (DQN). 
Numerical simulations validate the performance for our proposed joint optimization scheme. Comparative results demonstrate that the developed D3QN-PER-based algorithm outperforms existing state-of-the-art learning approaches which verifies its superiority in improving SEE for UAV-IRS-assisted LAE wireless communications network.
\end{abstract}

\section{INTRODUCTION}
\IEEEPARstart{A}{s} one of the key application scenarios for Low-Altitude Economy (LAE), urban environments feature complex communication conditions and stringent security requirements, which demand network infrastructures with ultra-low latency and high reliability. 
LAE concentrates on the development, utilization and management of low-altitude airspace, evolving into a collaborative ecosystem driven by diverse aerial platforms, such as unmanned aerial vehicles (UAVs)\cite{Networked_ISAC_for_Low-Altitude_Economy_Coordinated_Transmit_Beamforming_and_UAV_Trajectory_Design, Recent_Advances_in_Resource_Allocation_and_Beam_Prediction_for_Large_Language_Models_Empowered_ISAC_Systems}, electric vertical take-off and landing \cite{Networked_ISAC_based_UAV_Tracking_and_Handover_Towards_Low-Altitude_Economy}. 
Among these platforms, UAV has demonstrated extensive utilities, in numerous practical fields, i.e., modern logistics, emergency rescue, environmental monitoring, and urban construction. 
Especially in disaster relief scenarios, where ground communication infrastructures are often damaged or paralyzed, UAVs can rapidly establish flexible and reliable temporary communication links to support emergency rescue operations and provide seamless wireless coverage for remote disaster-stricken areas\cite{A_Beamforming-Aided_Full-Diversity_Scheme_for_Low-Altitude_Air-to-Ground_Communication_Systems_Operating_With_Limited_Feedback, A_Novel_Tensor_Decomposition-Based_Efficient_Detector_for_Low-Altitude_Aerial_Objects_With_Knowledge_Distillation_Scheme, UAV-Enabled_Multi-Pair_Massive_MIMO-NOMA_Relay_Systems_With_Low-Resolution_ADCs_DACs}.

Although UAVs possess distinct advantages in high mobility and flexible deployment, their air-to-ground communication links are easily blocked by complex terrain and surrounding obstacles in practical environments \cite{Energy_Efficient_Maximization_for_UAV-Mounted_RIS_Assisted_MEC_with_Backscatter_Systems, IRS-Assisted_Secure_UAV_Transmission_via_Joint_Trajectory_and_Beamforming_Design, When_UAV_Meets_IRS_Expanding_Air-Ground_Networks_via_Passive_Reflection, Reconfigurable_Intelligent_Surface_Equipped_UAV_in_Emergency_Wireless_Communications_A_New_FadingShadowing_Model_and_Performance_Analysis}. As a promising enabling technology for LAE communication systems, intelligent reflecting surface (IRS) exhibits great application potential in low-altitude internet of things scenarios \cite{Reflections_and_Prospects_of_Wireless_Coverage_Expansion_Technologies_for_Low-Altitude_Economy}. 
Specifically, IRS is a passive metamaterial array panel. 
This panel comprises numerous inexpensive reflective elements capable of independently tuning the amplitude, phase, polarization, and frequency of incident radio-frequency (RF) waves to enhance wireless link reliability \cite{Dynamic_Energy-Saving_Design_for_Double-Faced_Active_RIS-Assisted_Communications_With_Imperfect_CSI, Max-Min_Fairness_in_IRS-Aided_Multi-Cell_MISO_Systems_With_Joint_Transmit_and_Reflective_Beamforming,Aerial_Intelligent_Reflecting_Surface_Joint_Placement_and_Passive_Beamforming_Design_With_3D_Beam_Flattening,Intelligent_Reflecting_Surface_Assisted_Anti-Jamming_Communications_A_Fast_Reinforcement_Learning_Approach, QoS-Aware_Resource_Allocation_of_RIS-Aided_Multi-User_MISO_Wireless_Communications}. 
The IRS-assisted communication offers several key benefits, such as reduced energy consumption, reconfigurable propagation environments, interference suppression, extended coverage and flexible arrangement \cite{Low-Complexity_Channel_Estimation_and_Passive_Beamforming_for_RIS-Assisted_MIMO_Systems_Relying_on_Discrete_Phase_Shifts, Learning_to_Reflect_and_to_Beamform_for_Intelligent_Reflecting_Surface_With_Implicit_Channel_Estimation, Covert_Communications_With_Enhanced_Physical_Layer_Security_in_RIS-Assisted_Cooperative_Networks, Location-Aided_Distributed_Beamforming_for_Near-Field_Communications_With_Element-Wise_RIS}. 
Combining UAVs with IRSs not only provides a new degree of design freedom by enabling dynamic adjustment of IRS location with UAV mobility \cite{Optimization_of_Wireless_Relaying_With_Flexible_UAV-Borne_Reflecting_Surfaces}, but also expands communications coverage through the high maneuverability and cost-effective deployment of UAVs. 

However, in security-sensitive LAE scenarios, achieving effective coordination between UAV and IRS still faces many new challenges, such as eavesdropping risks \cite{Aerial-IRS-Assisted_Securing_Communications_Against_Eavesdropping_Joint_Trajectory_and_Resource_Allocation} and power consumption constrains. Although UAV assisted IRS communications can achieve panoramic/full-angle reflection, it may also enhance the communication performance of illegal users while improving the communication performance of legitimate users \cite{STAR-RIS-Assisted_Covert_Wireless_Communications_With_Randomly_Distributed_Blockages}. Thus, security issues have become critical research concerns for UAV-assisted IRS communications \cite{Safeguarding_Next-Generation_Multiple_Access_Using_Physical_Layer_Security_Techniques_A_Tutorial}. 
Inducing artificial noise—such as jamming UAVs \cite{A_Robust_Cooperative_Jamming_Scheme_for_Secure_UAV_Communication_via_Intelligent_Reflecting_Surface} and jamming beamforming \cite{ Secure_Transmission_for_IRS-on-UAV-assisted_Wireless_Networks}—along with adopting physical-layer security techniques \cite{Broadcast_Secrecy_Rate_Maximization_in_UAV-Empowered_IRS_Backscatter_Communications, Secure_Transmission_Design_for_Aerial_IRS_Assisted_Wireless_Networks} and designing covert communication schemes \cite{Covert_Communication_Assisted_by_UAV-IRS}, constitutes the primary approaches to address security challenges in UAV-assisted IRS communications networks. 
Exsiting studies deploy jamming UAVs as interference for eavesdroppers so as to guarantee the communications for legitimate users \cite{A_Robust_Cooperative_Jamming_Scheme_for_Secure_UAV_Communication_via_Intelligent_Reflecting_Surface} and generate artificial noise for covert communications. \cite{ Broadcast_Secrecy_Rate_Maximization_in_UAV-Empowered_IRS_Backscatter_Communications, Secure_Transmission_Design_for_Aerial_IRS_Assisted_Wireless_Networks}, some works maximize the secrecy capacity for UAV-RIS communications networks to ensure secure transmission. Other research \cite{Covert_Communication_Assisted_by_UAV-IRS, Secure_Transmission_for_IRS-on-UAV-assisted_Wireless_Networks} analyze the detection performance of eavesdroppers and maximize the covert transmission rate. 
However, most existing schemes adopt static IRS deployment on building surfaces or fixed UAV platforms, which limits flexible location optimization due to practical space and site constraints.

Moreover, as a passive device, the IRS can effectively reduce energy consumption in UAV assisted IRS wireless communications systems. Nevertheless, the limited endurance and propulsion power of UAV remain critical challenges that hinder practical deployment. To address this energy consumption issues in UAV assisted IRS communications, extensive research efforts have been dedicated in recent years. Specifically, researchers have employed alternating optimization algorithms \cite{IRS-UAV_Relaying_Networks_for_Spectrum_and_Energy_Efficiency_Maximization} and deep reinforcement learning (DRL) techniques \cite{Resource_Allocation_and_3D_Trajectory_Design_for_Power-Efficient_IRS-Assisted_UAV-NOMA_Communications, Learning-based_Energy_-_Efficiency_Optimization_for_IRS-assisted_Master-_Auxiliary-_Uav-_Enabled_Wireless-Powered_IoT_Networks}\cite{Active-IRS-Enabled_Energy-Efficiency_Optimizations_for_UAV-Based_6G_Mobile_Wireless_Networks, Intelligent_Surface-Assisted_UAV_Networks_A_DRL_Approach_to_Energy_Efficiency} to address these power-related challenges. 
In \cite{IRS-UAV_Relaying_Networks_for_Spectrum_and_Energy_Efficiency_Maximization}, an energy efficiency (EE) maximization scheme is designed under the strict on-board energy constraints of UAVs. By applying Dinkelbach’s method to resolve the non-convex fractional programming problem inherent in the trajectory optimization, the scheme achieves a rigorous trade-off between energy consumption and communication quality.
%
In \cite{Resource_Allocation_and_3D_Trajectory_Design_for_Power-Efficient_IRS-Assisted_UAV-NOMA_Communications}, the total energy consumption of wireless systems is minimized via a framework powered by deep neural networks (DNN). Specifically, the formulated optimization problem is partitioned into two separate problems, then addresses them alternately. 
Notably, in \cite{Learning-based_Energy_-_Efficiency_Optimization_for_IRS-assisted_Master-_Auxiliary-_Uav-_Enabled_Wireless-Powered_IoT_Networks}, a hybrid approach combining deep unsupervised learning (DUL) and artificial replay buffer initialization (ARBI) is proposed to enhance energy efficiency. 
In \cite{Active-IRS-Enabled_Energy-Efficiency_Optimizations_for_UAV-Based_6G_Mobile_Wireless_Networks}, a multi-objective hierarchical DRL-based energy minimization scheme is developed for the simultaneous reduction of power expenditure incurred by UAVs as well as ground users.
Similarly, \cite{Intelligent_Surface-Assisted_UAV_Networks_A_DRL_Approach_to_Energy_Efficiency} introduces a DRL-based strategy specifically designed to mitigate energy consumption in such systems. The above studies have made valuable contributions to improving secrecy performance or reducing power consumption in UAV-assisted IRS communications networks. However, the above-discussed studies concentrates on addressing only one of the two issues—either secrecy or power consumption—in secure UAV-IRS networks operating within the LAE scenario. 

Motivated by the aforementioned challenges, this paper studies the secrecy energy efficiency optimization concerns for UAV-assisted IRS wireless communication network, in which the base station (BS) conveys confidential data to legitimate users via signal reflection through IRSs mounted on UAVs, while an eavesdropper attempts to intercept the communication link. To ensure secure transmission for legitimate users and mitigate eavesdropping risks, we cooperatively optimize the BS transmit beamformer, IRS reflective units, and UAV trajectory to maximize the SEE in UAV-assisted IRS communication networks. To effectively solve the SEE maximization problem, we adopt the alternating optimization method and propose a D3QN-PER-based algorithm. We conduct numerical simulations to evaluate the performance of the proposed joint optimal solution, and perform comparative analyses between the D3QN-PER-based algorithm and other advanced learning methods. The following summarizes the main contributions of this paper.
\begin{itemize}
	\item We focus on the SEE optimization problem in LAE scenarios, where the critical application of integrating UAVs with IRSs is considered. UAV-assisted IRS communications can expand communications coverage, enhance flexible deployment, and mitigate blockage. The omnidirectional reflection characteristic renders the communication link vulnerable to eavesdropping. Also, the energy consumption of UAVs necessitates more efficient utilization of existing energy resources. Thus, SEE is a critical issue in LAE scenarios to be tackled with. 
\end{itemize}
\begin{itemize}
	\item To achieve secure, low-cost and flexible system design, we formulate the SEE maximization problem that optimizes the BS beamforming vector, IRS phase shift, and UAV trajectory. To address this problem, we adopt the alternative optimization method, whereby the SEE maximization problem is decomposed into three subproblems, i.e., beamforming vector optimization, IRS phase shift optimization, and UAV trajectory optimization. To make the non-convex fractional structure tractable, we employ Dinkelbach’s method to convert it into a computationally subtractive form. Then, introducing slack variables and utilizing Semidefinite Relaxation (SDR). 
Through this systematic approach, we compute the optimal beamforming vector and the optimal phase shift vector within the UAV-assisted IRS communications networks.
\end{itemize}
\begin{itemize}
	\item Although DQN is easy to implement, it suffers from slow and unstable convergence in practical training. To overcome this limitation, we integrate the advantages of three advanced improved schemes, i.e., the Dueling DQN architecture, the Double DQN (DDQN) mechanism, and the DDQN framework integrated with Prioritized Experience Replay (DDQN-PER). On this basis, we develop the D3QN-PER algorithm to settle the independent UAV trajectory subproblem. From simulation measurements, our designed framework achieves obvious performance improvement when contrasted with other benchmark methods.
\end{itemize}

The rest of the paper is organized as follows. Section II introduces the channel and transmission model of the UAV-IRS assisted LAE communications system. Section III proposes SEE maximization problem. Section IV provides the proposed algorithms. Section V evaluates the performance of our proposed designs. The paper concludes with Section VI.

$Notations$: We use bold uppercase letters for matrices and bold lowercase letters for vectors. $\mathbb{C}^{N \times M}$ is the space of $N$ × $M$. $\mathbb{E}\{ \bullet \}$ represents the statistical expectation, and Re$\{ \bullet \}$ denotes the real part of a complex number. For a square matrix $\mathbf{A}$, Tr$({\bf{A}})$, rank$({\bf{A}})$ and $\mathbf{\mathbf{A}} \succcurlyeq 0$ represent trace, rank, and positive semi-defifinite of $\mathbf{A}$, respectively. For a vector, $\left| \bullet \right|$ represents its Euclidean norms. For a complex number, $\left\| \bullet \right\|$ is its magnitude. The ${( \bullet )^\mathrm{H}}$ and ${( \bullet )^ \mathrm{T} }$ denote Hermitian operation and transpose, respectively. 
diag$({\bf{A}})$ represents a diagonal matrix constructed with vector $\mathbf{a}$'s elements on its main diagonal.
\section{Systerm Model}
We consider a UAV-assisted IRS securing wireless communications network relying on IRS deployed on aerial platforms aided by UAVs, as shown in Fig. 1. 
The communication signals are delivered from the BS to $K$ legitimate users via the direct link as well as the reflected link facilitated by the IRS. On the ground, a potential eavesdropper (Eve) is placed with the objective of intercepting the communications between the BS and users. 
To facilitate geometric modeling of the network, we model the network in a 3D Cartesian coordinate system, wherein the spatial coordinates of the BS, Eve, UAV and the $k$th ($k \in K$) user are specified as $(0, 0, 0)$, $(x_{e}, y_{e}, 0)$, $(x_{a},y_{a},z_{a})$ and $(x_{t,k} ,0,0)$,  respectively, with coordinates in meters.
\begin{figure}[t]
    \centering
    \includegraphics[scale=0.5, trim=60pt 60pt 20pt 40pt, clip]{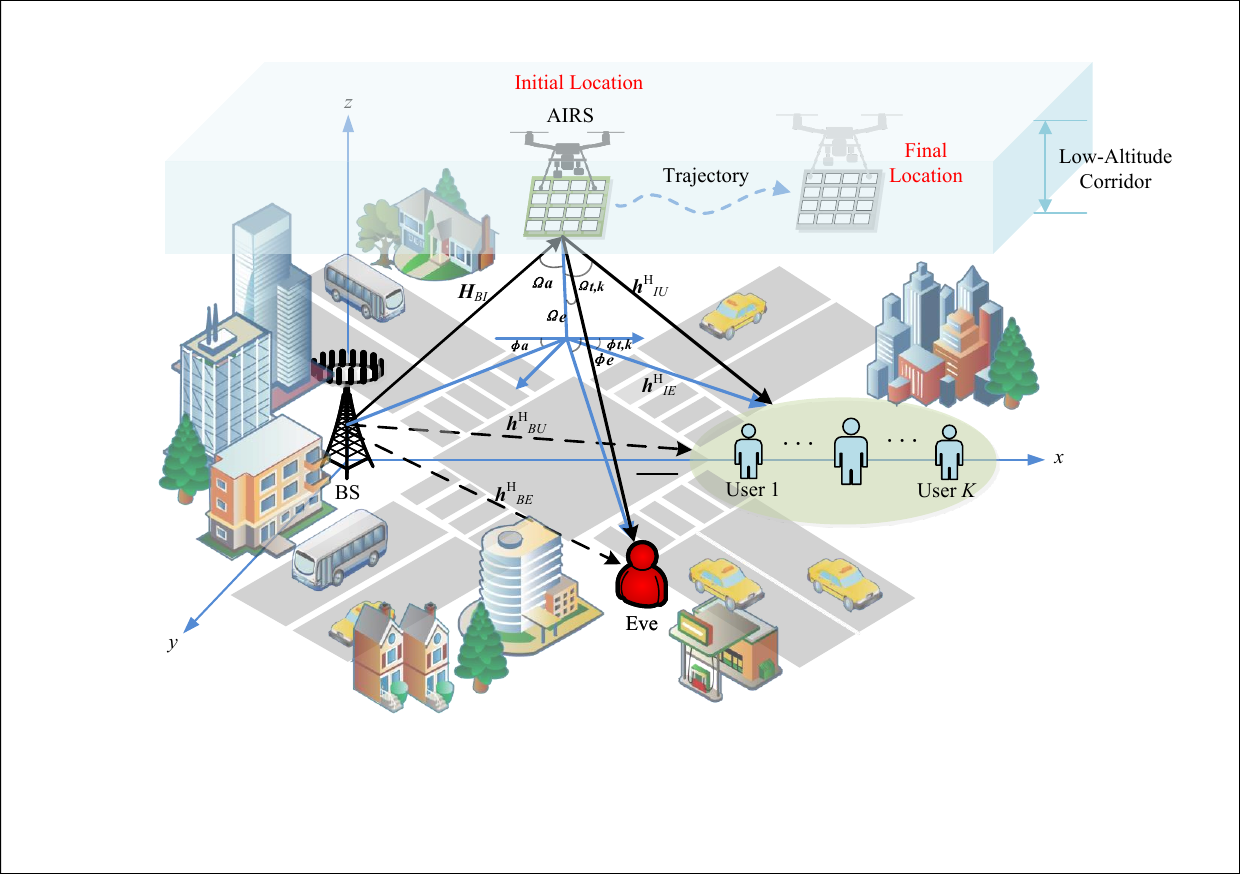}
    \caption{The UAV-IRS securing wireless communications.}
\end{figure}
\subsection{Channel Model}
We assume that there are $M$ antennas on the BS, while each user and Eve are configured with single-antenna. IRSs are flat surfaces consisting of $N$ passive reflective units (where $N=N_{x} N_{y}$). Driven by its integrated intelligent controller, IRS can dynamically adjust each reflective element's phase shift. $\mathbf{H}_{BI} \in \mathbb{C}^{N \times M}$, $\mathbf{h}_{BU,k}^{H} \in \mathbb{C}^{1 \times M}$ and $\mathbf{h}_{BE}^{H} \in \mathbb{C}^{1 \times M}$ correspond to the channel gains of BS$\to$IRS, BS$\to$user $k$ and BS$\to$Eve propagation paths, respectively. Let $\mathbf{h}_{IU,k}^{H} \in \mathbb{C}^{1 \times N}$, $\mathbf{h}_{IE}^{H} \in \mathbb{C}^{1 \times N}$ be the channel gains of IRS$\to$user $k$ and IRS$\to$Eve propagation paths, respectively. We denote $\mathbf{\Theta} = \mathrm{diag}\left( {\beta e^{j\theta_{1}},\beta e^{j\theta_{2}},\ldots,\beta e^{j\theta_{N}}} \right)$ as the IRS's phase-shift matrix. To characterize the fundamental performance units of IRS, we assume that the reflection unit phase angles, denoted by  $\theta_n\in\left[0,\right.\left.2\pi\right)$ $(n=1,2,\ldots,N)$, are continuous. To prevent signal transmission through the IRS and the resultant attenuation of desired communication signals, we assume full-reflection configuration for IRS in our analysis. Thus, each reflective component's amplitude is fixed at its maxinum, i.e., $\beta =1$.

The flying trajectory of UAV is denoted as $q=(x_{a},y_{a},z_{a})$ in the 3D Cartesian coordinate. Let $d_{\mathrm {BI}}$, $d_{\mathrm {IU},k}$ and $d_{\mathrm {IE}}$ denote the distances for BS$\to$IRS, IRS$\to$user $k$, and IRS$\to$Eve, respectively. Accordingly, $d_{\mathrm {BI}} $, $d_{\mathrm {IU},k}$ and $d_{\mathrm {IE}}$ can be calculated using $ \sqrt{x_{a}^{2}+y_{a}^{2}+z_{a}^{2}}$, $ \sqrt{\left (x_{a}-x_{t,k}\right)^{2}+y_{a}^{2}+z_{a}^{2}}$ and $ \sqrt{\left (x_{a}-x_{e} \right)^{2}+\left (y_{a}-y_{e} \right)^{2}+z_{a}^{2}}$, respectively. Then, the channel power gains for BS$\to$IRS, IRS$\to$user $k$, and IRS$\to$Eve, denoted by $\beta _{\mathrm{BI}}\left ( q \right ) $, $\beta _{\mathrm{IU,k} }\left ( q \right )$ and  $\beta _{\mathrm{IE}}\left ( q \right ) $, respectively, can be expressed as follows
	\begin{eqnarray}
		\begin{cases}
			\begin{aligned}
				&\beta _{\mathrm{BI} }\left ( q \right ) =\beta _{0}d_{\mathrm {BI}}^{-2}=\frac{\beta_{0}}{\sqrt{x_{a}^{2}+y_{a}^{2}+z_{a}^{2}}},
				\\
				&\beta _{\mathrm{IU},k}\left ( q \right )  =\beta _{0}d_{\mathrm {IU},k}^{-2}=\frac{\beta_{0}}{\sqrt{\left (x_{a}-x_{t,k}\right)^{2}+y_{a}^{2}+z_{a}^{2}}},
				\\
				&\beta _{\mathrm{IE} }\left ( q \right ) \!=\! \beta _{0}d_{\mathrm {IE}}^{-2}\!=\!\frac{\beta_{0}}{\sqrt{\left (x_{a}\!-\!x_{e} \right)^{2}\!+\!\left (y_{a}\!-\!y_{e} \right)^{2}+z_{a}^{2}}},
			\end{aligned}
		\end{cases}
	\end{eqnarray}
where $\beta _{0}$ is the channel power gain corresponding to the benchmark distance of 1 m.

Assume that the wireless channels operate under both Line of Sight (LOS) and Non Line of Sight (NLOS) propagation environments. Under such channel conditions, we choose the Rice channel model. Then, the channel matrix for BS$\to$IRS can be expressed as follows
	\begin{eqnarray}
		\mathbf{H}_{BI}\left(q\right) \!=\! \sqrt{\beta_{BI}(q)} \left (\sqrt{\frac{K_1}{K_1\!+\!1}}\psi _{a} \left ( q \right ) \!+\! \sqrt{\frac{1}{K_1 \!+\! 1}}\mathbf{H}_{\text{N}} \right ),
		\label{eq:3}
	\end{eqnarray}
where $ K_1 $ denotes the Rice factor of the base station to the IRS channel, $\psi _{a} \left ( q \right ) \in \mathbb {C}^{N\times M}$ represents the channel matrix associated with the direct link, while $\mathbf{H}_{\text{N}}\in \mathbb {C}^{N\times M}$ denotes the multipath scattering channel matrix. We denote the multipath scattering channel matrix linking the BS and IRS as $\mathbf{H}_{\text{N}} \!=\! \left ( \tau_{x}  + j\tau_{y} \right )/\sqrt{2}$, where $\tau_{l}\sim \mathcal{ CN}\left ( 0,1 \right )$, $l\in\left \{ x,y \right \} $.

We define $\phi_{a}\left ( q \right)$ as the azimuth angle of arrival (AoA) for the BS$\to$IRS signal, representing the angle between the wave propagation projected onto the $xoy$ plane and the $x$-axis. Similarly, $\phi_{u,k}\left ( q \right)$ and $\phi_{e}\left ( q \right)$ denote the azimuth Angles of Departure (AoD) for  the IRS$\to$user $k$ and IRS$\to$Eve link, respectively, measured from the $x$-axis within the $xoy$ plane. Additionally, $\phi_{b}\left ( q \right)$ is defined as the AoD over the BS$\to$IRS link. Furthermore, the elevation angles for the BS$\to$IRS, IRS$\to$user $k$, and IRS$\to$Eve channels are denoted by $\Omega_{a}\left ( q \right)$, $\Omega_{u,k}\left ( q \right)$ and $\Omega_{e}\left ( q \right)$, respectively, which quantify the inclination of the wave propagation from the positive $z$-axis. Consequently, the direct propagation channel matrix linking the BS and IRS is represented as follows
	\begin{eqnarray}
		\psi _{a} \left ( q \right )   =\mathbf{a}_{\text{IRS}}(\Omega_{a}\left ( q \right ) , \phi_{a}\left ( q \right ) )\mathbf{a}_{\text{BS}}^H(\phi_{b}\left ( q \right)),
	\end{eqnarray}
 where $\mathbf{a}_{\text{IRS}}$ denotes the receive array response vector for the AoAs over the BS$\to$IRS link, and $\mathbf{a}_{\text{BS}}(\phi_{b}\left ( q \right ))$ represents the transmit array response vector corresponding to the AoDs of this same propagation path. Let us denote $\bar{u}_{a}  = \sin\Omega_{a}\left ( q \right )  \cos\phi_{a}\left ( q \right ) $, $\bar{v}_{a}  = \sin\Omega_{a}\left ( q \right )  \sin\phi_{a}\left ( q \right ) $, $\bar{u}_{u, k}  = \sin\Omega_{u,k}\left ( q \right )  \cos\phi_{u,k}\left ( q \right ) $, $\bar{v}_{u,k}  = \sin\Omega_{u,k}\left ( q \right )  \sin\phi_{u,k}\left ( q \right ) $, and $\bar{u}_{e}  = \sin\Omega_{e}\left ( q \right )  \cos\phi_{e}\left ( q \right ) $, $\bar{v}_{e}  = \sin\Omega_{e}\left ( q \right )  \sin\phi_{e}\left ( q \right ) $. Here, $\sin\Omega_{a}\left ( q \right ) \!=\!\sqrt{x_{a}^{2} + y_{a}^{2}} / d_{\mathrm{BI}}$, $\sin\phi_{a}\left ( q \right ) = y_{a}/ \sqrt{x_{a}^{2} + y_{a}^{2}} $, $\cos\phi_{a}\left ( q \right ) = x_{a}/ \sqrt{x_{a}^{2} + y_{a}^{2}} $. The received array response matrix and the transmit array response matrix are formulated as follows
	\begin{eqnarray}
		\begin{cases}
			\begin{aligned}
				&\mathbf{a}_X(\bar{u}_{a} ) = \left[1, e^{-j\frac{2\pi d}{\lambda _{\mathbf{a}} } \bar{u}_{a}  }, \dots, e^{-j\frac{2\pi d}{\lambda _{\mathbf{a}} } (N_X-1) \bar{u}_{a} } \right],
				\\
				&
		\mathbf{a}_Y(\bar{v}_{a} ) = \left[1, e^{-j\frac{2\pi d}{\lambda _{\mathbf{a}} } \bar{v}_{a} }, \dots, e^{-j\frac{2\pi d}{\lambda _{\mathbf{a}} } (N_Y-1) \bar{v}_{a}} \right],
			\end{aligned}
		\end{cases}
	\end{eqnarray}
and
\begin{eqnarray}
	\begin{cases}
		\begin{aligned}
			&\mathbf{a}_{\text{IRS}}(\Omega_{a }\left ( q \right ), \phi_{a}\left ( q \right ))\! =\! \mathbf{a}_X(\bar{u}_{a} ) \otimes \mathbf{a}_Y(\bar{v}_{a}  ),
			\\
			&
	\mathbf{a}_{\text{BS}}(\phi_{b}\left ( q \right ))\!\! =\!\! \left[\!1,\! e^{-j\frac{2\pi d}{\lambda _{\mathbf{a}} }   sin\phi_{b}}, \dots, e^{-j\frac{2\pi d}{\lambda _{\mathbf{a}} } (M\!-\!1) sin\phi_{b} }\! \right]\!\!,
		\end{aligned}
	\end{cases}
\end{eqnarray}
where  $\otimes$ denotes Hadamard inner products, $\mathbf{a}_X$ is the array response vector in the horizontal direction of the IRS, and $\mathbf{a}_Y$ denotes the array response vector in the vertical direction of the IRS. In Eq.~(5), $\lambda _{\mathbf{a}} $ is the carrier wavelength and $d$ is the antenna separation. We can calculate the channel matrices for the direct paths IRS$\to$user $k$ and IRS$\to$Eve in the same way as for $\psi _{a} \left ( q \right )$, denoted by $\psi _{u, k} \left ( q \right )$ and $\psi _{e} \left ( q \right )$, respectively, can be expressed as follows
\begin{eqnarray}
	\begin{cases}
		\begin{aligned}
			&\psi _{u, k} \left ( q \right ) = \mathbf{a}_X(\bar{u}_{u,k} ) \otimes \mathbf{a}_Y(\bar{v}_{u,k}  ),
			\\
			&\psi _{e} \left ( q \right ) = \mathbf{a}_X(\bar{u}_{e} ) \otimes \mathbf{a}_Y(\bar{v}_{e}  ).
		\end{aligned}
	\end{cases}
\end{eqnarray}
Then, the channel matrix $\mathbf{h}_{IU,k}\left ( q \right)$ and $\mathbf{h}_{IE}\left ( q \right)$ can be written as follows
\begin{eqnarray}
	\begin{cases}
		\begin{aligned}
			&\!\mathbf{h}_{IU,k}\!\left ( q \right)\!\! = \!\!\sqrt{\!\beta_{IU,k}}\! \left (\!\sqrt{\!\frac{K_1}{K_1\!+\!1}} \psi _{u,k} \!+\! \sqrt{\!\frac{1}{K_1\!+\!1}} \!\mathbf{H}_{\text{N}}\! \right ),
			\\
			&\!\mathbf{h}_{IE}\left ( q \right) \!\!=\! \sqrt{\beta_{IE}} \left (\sqrt{\frac{K_1}{K_1\!+\!1}}\!\psi _{e}\!+\! \sqrt{\frac{1}{K_1\!+\!1}} \mathbf{H}_{\text{N}} \right ).
		\end{aligned}
	\end{cases}
\end{eqnarray}

When the position of the IRS changes, channel parameters will change accordingly. In order to suppress the interference from illegitimate users and at the same time enhance received information of legitimate users, it is necessary to compute the received signals of all users on the basis of the updated channel matrix.

\subsection{Transmission Model}
We assume that $\mathbf{x}$ is the transmit signal by the BS. Here $\mathbf{x} = \mathbf{f}s$, where $s \sim \mathcal{C}\mathcal{N}\left( 0,1 \right)$ and $\mathbf{f} \in \mathbb{C}^{M \times 1}$ denote the independent and identically distributed complex Gaussian random variables and the beamforming vector, respectively. Let $P_\mathrm{max}$ represent the upper bound of the BS transmit power, where the transmit beamforming vector $\mathbf{f}$ satisfies $\mathbf{f}^{H}\mathbf{f} \leq P_\mathrm{max}$. 
Based on the system model in \cite{Weighted_Sum-Rate_Maximization_for_Reconfigurable_Intelligent_Surface_Aided_Wireless_Networks}, let $y_{u,k}$ and $y_{e}$ denote the received signals at the $k$th user and Eve, which can be expressed as follows
\begin{eqnarray}
	\begin{cases}
		\begin{aligned}
			&y_{u,k} = \left( \mathbf{h}_{Bu,k}^{H}  + \mathbf{h}_{Iu,k}^{H}\left ( q \right ) \mathbf{\Theta} \mathbf{H}_{BI}\left ( q \right )  \right)\mathbf{f}s + n_{u},
			\\
			&y_{e} = \left( \mathbf{h}_{BE}^{H} + \mathbf{h}_{IE}^{H }\left( q \right )\mathbf{\Theta} \mathbf{H}_{BI}\left( q \right ) \right)\mathbf{f}s + n_{e},
		\end{aligned}
	\end{cases}
\end{eqnarray}
where $n_{u}$ and $n_{e}$ represent the complex additive white Gaussian noise (AWGN) at the user and Eve, respectively. Here  $\mathcal{C}\mathcal{N}\left( 0,\sigma_{u}^{2} \right)$ and $\mathcal{C}\mathcal{N}\left( 0,\sigma_{e}^{2} \right)$ characterize the distributions of zero-mean complex Gaussian random variables with variances $\sigma_{u}^{2} $, $\sigma_{e}^{2} $, respectively. 
Based on the system model in  \cite{Joint_Power_Control_and_Passive_Beamforming_in_IRS-Assisted_Spectrum_Sharing,Intelligent_Reflecting_Surface_Enabled_Secure_Cooperative_Transmission_for_Satellite-Terrestrial_Integrated_Networks}, let $\gamma_{u,k}$ and $\gamma_{e}$ denote the received SNRs for the $k$th user and Eve, respectively, which are calculated as follows
\begin{eqnarray}
	\begin{cases}
		\begin{aligned}
			&\gamma_{u,k} = \frac{\left| {\left. \left( \mathbf{h} \right._{Bu,k}^{H} + \mathbf{h}_{Iu,k}^{H}\left ( q \right )\mathbf{\Theta} \mathbf{H}_{BI}\left ( q \right ) \right)\mathbf{f}} \right|^{2}}{\sigma_{u}^{2}},
			
			\\
			&\gamma_{e} = \frac{\left| {\left( \mathbf{h}_{BE}^{H} + \mathbf{h}_{IE}^{H}\left ( q \right )\mathbf{\Theta} \mathbf{H}_{BI} \left ( q \right )\right)\mathbf{f}} \right|^{2}}{\sigma_{e}^{2}},
		\end{aligned}
	\end{cases}
\end{eqnarray}

Let us denote $R_{u,k}$ and $R_{e}$ the channel transmission rates of the user $k$, and the channel transmission rate of Eve, respectively, which are written as follows
\begin{eqnarray}
	\begin{cases}
		&R_{u,k} = B{\log_{2}\left( {1 + \gamma_{u,k}} \right)},
		\\
		&R_{e} = B{\log_{2}\left( {1 + \gamma_{e}} \right)},
	\end{cases}
\end{eqnarray}
where $B$ represents the transmission channel bandwidth. 

To evaluate the secrecy level of multi-user IRS-assisted wireless communication systems, we take the minimum achievable secrecy rate over all legitimate users as the core performance indicator. According to \cite{Intelligent_Reflecting_Surface_Aided_Multi-Antenna_Secure_Transmission}, the minimum achievable secrecy rate over the UAV-IRS assisted LAE communications system can be expressed as follows
\begin{equation}
	\begin{aligned}
		R_\mathrm{s} = \min\limits_{k \in K}{B{\log_{2}{\left( {1 + \gamma_{u,k}} \right) - B{\log_{2}\left( {1 + ~\gamma_{e}} \right)}}}}.
	\end{aligned}
\end{equation}

Assume that $P_\mathrm{BS}$ and $P_\mathrm{IRS}$ represent the power consumption at the multi-antenna BS and IRS, respectively. $\xi$ denotes the amplification coefficient preceding the transmit beamforming vector. Thus, the overall power consumption across the LAE-assisted secrecy wireless communication system, denoted by $P_\mathrm{tot}$, can be expressed as follows
\begin{eqnarray}
	\begin{aligned}
		P_\mathrm{tot} = \xi\left( {\mathbf{f}^{H}\mathbf{f}} \right) + P_\mathrm{BS} + P_\mathrm{IRS},
	\end{aligned}
\end{eqnarray}
Then, the energy efficiency for the security aware UAV assisted IRS communications networks is defined as follows
\begin{eqnarray}
	\eta_{se} = \frac{\min\limits_{k \in K}{B{\log_{2}{\left( {1 + \gamma_{u,k}} \right) - B{\log_{2}\left( {1 + ~\gamma_{e}} \right)}}}}}{P_\mathrm{tot}}.
\end{eqnarray}

The beamforming at BS, reflection coefficients of the IRS, and UAV trajectory are all key factors affecting the secrecy energy efficiency. To maximize the SEE $\eta_{se}$, we jointly optimize these three coupled variables.
\section{Secrecy  energy efficiency  Maximization Problem for UAV-IRS Communications networks}\label{sec:pf}
In this section, we formulate the SEE maximization problem for the UAV-IRS-assisted LAE wireless communications networks. Because the fractional structure of SEE introduces calculation obstacles, we leverage Dinkelbach’s method to convert the formula into an auxiliary parametric subtraction.
\subsection{Problem Formation}
We aim at maximizing the SEE for UAV-assisted IRS communications networks while adhering to the upper-bound power restriction and the target secrecy rate requirement. We formulate the SEE maximization problem, denoted by $\textbf{P0}$, which is shown as follows
\begin{equation*}
    \begin{split}
        \textbf{P0:}~&\mathop {\max\limits_{\mathbf{f},\theta,q}\frac{~\min\limits_{k \in K}{B{\log_{2}{\left( {1 + \gamma_{u,k}} \right) - B{\log_{2}\left( {1 + ~\gamma_{e}} \right)}}}}}{\xi\left( {\mathbf{f}^{H}\mathbf{f}} \right) + P_\mathrm{BS} + P_\mathrm{IRS}}}\\
        \mathrm{s.t.}\quad &\begin{cases}
            C1:\mathbf{f}^{H}\mathbf{f} \leq P_\mathrm{max},\\
            C2:\min\limits_{k \in K}\left\{ B{\log_{2}\! \!{\left. \left( {1 \!\! + \! \!\gamma_{u,k}} \right) \right\} - B{\log_{2}\! \!{\left( {1 \! \! + \! \!\! ~\gamma_{e}} \right) \geq {R_\mathrm{min}} }}}} \right.,\\
            C3:\theta_{n} \in \left\lbrack {0,\left. {2\pi} \right)} \right.,~n \in 1,2,\ldots,N,\\
        \end{cases}
    \end{split}
\end{equation*}
where $R_\mathrm{min}$ is the minimum achievable secrecy rate. Obviously, the objective function and constraint C2 are non-convex. Thus, SEE maximization problem $\textbf{P0}$ is non-convex. To solve for $\textbf{P0}$, we first employ Dinkelbach’s method to simplify the calculation.
\subsection{Dinkelbach’s method}
For simplicity, according to \cite{Energy_Efficiency_in_Secure_IRS-Aided_SWIPT}, we employ a non-negative auxiliary variable $\lambda$ to transform the fractional objective function into a subtractive form—a method known as the Dinkelbach. Then, problem P0 can be reformulated as problem P1, shown as follows
\begin{equation*}
	\begin{split}
		\textbf{P1:}&\mathop {{\max\limits_{f,\theta, q}~}{\underset{k \in K}{\left\{ \min \right.}{B{\log_{2}\!\!{\left. \left( {1 \!\! + \!\! \gamma_{u,k}} \right) \right\} - B{\log_{2}\left( {1 \!\! + \!\! ~\gamma_{e}} \right)}}}}}} - \lambda P_{tot}\\
		\mathrm{s.t.}\quad &\begin{cases}
        C1:\mathbf{f}^{H}\mathbf{f} \leq P_\mathrm{max},\\
		C2:\min\limits_{k \in K}\left\{ B{\log_{2}\! \!{\left. \left( {1 \!\! + \! \!\gamma_{u,k}} \right) \right\} - B{\log_{2}\! \!{\left( {1 \! \! + \! \!\! ~\gamma_{e}} \right) \geq {R_\mathrm{min}} }}}} \right.,\\
		C3:\theta_{n} \in \left\lbrack {0,\left. {2\pi} \right)} \right.,~n \in 1,2,\ldots,N,\\
      \end{cases}
	\end{split}
\end{equation*}

Let us define $\mathcal{\mathbf{\omega }}_{N}\mathord{=}e^{j\theta_{N}}$ and 
$\mathcal{\mathbf{w}}^{H}\mathord{=}\left( {{\omega }_{1},{\omega }_{2},\ldots,{\omega }_{N}} \right)$. Then, $\mathbf{h}_{Iu,k}^{H}\left ( q \right ) \mathbf{\Theta} \mathbf{H}_{BI}\left ( q \right )$, and $\mathbf{h}_{IE}^{H}\left ( q \right )\mathbf{\Theta} \mathbf{H}_{BI}\left ( q \right )$ are formulated as follows 
\begin{eqnarray}
	\begin{cases}
		&\mathbf{h}_{Iu,k}^{H}\left ( q \right ) \mathbf{\Theta} \mathbf{H}_{BI}\left ( q \right ) =\mathcal{\mathbf{w}}^{H}\mathbf{H}_{IBU}\left ( q \right ) ,\\
		
		&\mathbf{h}_{IE}^{H}\left ( q \right )\mathbf{\Theta} \mathbf{H}_{BI}\left ( q \right )=\mathcal{\mathbf{w}}^{H}\mathbf{H}_{IBE}\left ( q \right ) ,
	\end{cases}
\end{eqnarray} 
where $\mathbf{H}_{IBU}\left ( q \right ) =\mathrm{diag}\left( \mathbf{h}_{Iu,k}^{H}\left ( q \right )  \right)\mathbf{H}_{BI}\left ( q \right ) $, $\mathbf{H}_{IBE}\left ( q \right )  = \mathrm{diag}\left( \mathbf{h}_{IE}^{H} \left ( q \right ) \right)\mathbf{H}_{BI}\left ( q \right ) $. Then, the equivalently transformed received SNR at the $k$th user and Eve, denoted by $\gamma_{u,k}'$ and $\gamma_{e}'$, respectively, can be rewritten as follows
\begin{eqnarray}
	\begin{cases}
		&\gamma_{u,k}'  = a_{1}\left| {{\overline{\mathcal{\mathbf{w}}}}^{H}\mathbf{H}_{1}\left ( q \right ) \mathbf{f}} \right|^{2},\\
		
		&\gamma_{e}'  = a_{2}\left| {{\overline{\mathcal{\mathbf{w}}}}^{H}\mathbf{H}_{2}\left ( q \right ) \mathbf{f}} \right|^{2},
	\end{cases}
\end{eqnarray} 
where $a_{1}\!\!=\!\! 1/\sigma_{u}^{2}$, $a_{2}\!= \!1/\sigma_{e}^{2}$, $\mathbf{H}_{1}\left ( q \right )\!\!=\!\!\left ( \mathbf{H}_{IBU}^{H}\left ( q \right ), \mathbf{h}_{Bu,k}\right )^{H}$ and $\mathbf{H}_{2}\left ( q \right )\!\!=\!\!\left ( \mathbf{H}_{IBE}^{H}\left ( q \right ), \mathbf{h}_{BE}\right )^{H}$. As specified in Eq.~(15), $ {\overline{\mathcal{\mathbf{w}}}}^{H} \!\!\!=\!\! e^{j{\mathfrak{w}}}\left (\mathbf{w}^{H},1\right) $, with $\mathfrak{w}$ denoting an arbitrarily chosen phase rotation parameter. The problem $\textbf{P1}$ can be converted to problem $\textbf{P2}$, shown as follows
\begin{equation*}
	\begin{split}
		\textbf{P2:}~&\mathop {\max\limits_{\mathbf{f},\overline{\mathcal{\mathbf{w}}}, q}{\underset{k \in K}{~\min\{}{\frac{B}{\ln 2}{\ln_{}\!\!{\left. \left( {1 \!\!+ \!\!\gamma_{u,k}' } \right) \right\} \!-\! \frac{B}{\ln 2}{\ln_{}\!\!\left( {1 \!\! + \!\! ~\gamma_{e}' } \right)}}}}}} \! - \! \lambda P_{tot}\\
		\mathrm{s.t.}\quad &\begin{cases}
        C1:\mathbf{f}^{H}\mathbf{f} \leq {P_\mathrm{max}},\\
		C2':{\underset{k \in K}{\min\{}{\frac{B}{\ln 2}{\ln_{}\!\!{\left. \left( {1 \!\! + \!\! \gamma_{u,k}' } \right) \right\} - \frac{B}{\ln 2}{\ln_{}\!\!\left( {1 \!\! + \!\! ~\gamma_{e}}'  \right)}}}}} \!\!\geq\!\! {R_\mathrm{min}},\\
		C3':\left| {\omega }_{n} \right| = 1,~n \in 1,2,\ldots,N.\\
     \end{cases}
	\end{split}
\end{equation*}

To solve this problem for the security aware UAV-assisted IRS communications networks, we decompose it into three subproblems, i.e., beamforming vector optimization subproblem, reflection phase shift optimization subproblem and UAV trajectory optimization subproblem. The three subproblems are alternately optimized.

\section{Joint Optimization for beamforming, IRS phase shift and UAV trajectory}\label{sec:pf}

To solve the SEE maximization problem for UAV-IRS-assisted wireless communications network, we focus on this high-dimensional problem dominated by three variables $\mathbf{f}$, $\overline{\mathcal{\mathbf{w}}}$, and ${q}$, which are optimized alternatively.

\subsection{Optimizing $\mathbf{f}$ with Given $\overline{\mathcal{\mathbf{w}}}$ and ${q}$}
With $\overline{\mathcal{\mathbf{w}}}$ and ${q}$ fixed, we formulate a single-variable subproblem and solve it by optimizing transmit beamforming vector $\mathbf{f}$ to maximize the SEE.

Let us denote ${\overline {\mathbf{h}_{1}\left ( q \right ) }}^{H} = {{\overline{\mathcal{\mathbf{w}}}}^{H}}{\mathbf{H}_{1}\left ( q \right )},{\overline {\mathbf{h}_{2}\left ( q \right ) }}^{H} = {{\overline{\mathcal{\mathbf{w}}}}^{H}}{\mathbf{H}_{2}\left ( q \right )}$, we have $\mathbf{F} = \mathbf{f}\mathbf{f}^{H}$, $\mathbf{F} \succcurlyeq 0$ and $\mathrm{rank}\left( \mathbf{F} \right) = 1$. Thus, we can obtain
\begin{eqnarray}
	\begin{cases}
		&\left| {{\overline{\mathbf{h}_{1}\left (q\right ) }}^{H}\mathbf{f}} \right|^{2} =  \mathrm{Tr}\left(\overline{ \mathbf{H}_{{i}}\left(q\right)} \mathbf{F} \right),\\
		
		&\left| {{\overline{\mathbf{h}_{2}\left (q\right ) }}^{H}\mathbf{f}} \right|^{2} =  \mathrm{Tr}\left(\overline{ \mathbf{H}_{{j}}\left(q\right)} \mathbf{F} \right),
	\end{cases}
\end{eqnarray} 
Where ${\overline{\mathbf{H}_{i}\left(q\right)} = {\overline{\mathbf{h}_{1}\left (q\right)}} \ {\overline{\mathbf{h}_{1}\left(q\right)}}^{\textit{H}}}$ and ${\overline{\mathbf{H}_{j}\left(q\right)} = {\overline{\mathbf{h}_{2}\left (q\right)}} \ {\overline{\mathbf{h}_{2}\left(q\right)}}^{\textit{H}}}$. Then, the problem $\textbf{P2}$ can be transformed into problem $\textbf{P3}$ using Eq.~(16), shown as follows
\begin{equation*}
	\begin{split}
		\textbf{P3: \ }&\max_{\mathrm {f}}~\{\min_{k\in K}\{\frac{B}{\ln{2}}\ln_{}{(1+a_{1}\mathrm{Tr}(\overline{\mathbf{H}_{i}\left ( q \right )} \mathbf{F} )
  \}\}}\\ & ~~~~- \frac{B}{\ln 2}{\mathrm{ln}\left( {1 + ~a_{2}\mathrm{Tr}\left( {\overline{\mathbf{H}_{j}\left ( q \right ) }\mathbf{F}} \right)} \right)} - \lambda P_\mathrm{tot}\\
		\mathrm{s.t.}\quad &\begin{cases}
        C1':\mathbf{F} \in \mathcal{F},\\
		\tilde{C}2:{~\{}{\min\limits_{k \in K}\{ \frac{B}{\ln 2}{\ln_{}{( {1 + a_{1}\mathrm{Tr}( {{\overline{\mathbf{H}_{i}\left ( q \right )}} \mathbf{F}})})\}\}}}}\\ 
        ~~~~- \frac{B}{\ln 2}{\mathrm{ln}\left( {1 + ~a_{2}\mathrm{Tr}\left( {\overline{\mathbf{H}_{j}\left ( q \right ) }\mathbf{F}} \right)} \right)} \geq  {R_\mathrm{min}}.
        \end{cases}
	\end{split}
\end{equation*}

\textbf{Definition~1: \ }For any positive scalar $x > 0$, consider the function $\varphi\left( t \right) = - tx + {\ln t} + 1$, It can be verified that
\begin{eqnarray}\label{xindaoxuanze}
	\max\limits_{t > 0}{\varphi\left( t \right) = - {\ln x}}.
\end{eqnarray}
where the maximum is achieved when $t = 1/x$. 

According to the Definition~1, we set ${t = t}_{u,k}$, $x = 1 + ~a_{1}\mathrm{\mathrm{Tr}}\left( {\overline{\mathbf{H}_{i}\left ( q \right ) }\mathbf{F}} \right)$. Thus, we have
\begin{eqnarray}
	\begin{aligned}
		R_{u,k}\frac{\ln 2}{B} & = {\mathrm{ln}\left( {1 + ~a_{1}\mathrm {Tr}\left( {\overline{\mathbf{H}_{i}\left ( q \right ) }\mathbf{F}} \right)} \right)}\\
		& = \underset{t_{u,k} > 0}{\mathrm{min}}~\varphi_{u,k}\left( \mathbf{F},t_{u,k} \right),
	\end{aligned}
\end{eqnarray}
where  
\begin{eqnarray}
		{\varphi_{u,k}\left( \mathbf{F},t_{u,k} \right)\!\! =\!\! t}_{u,k}\!*\!\left\lbrack {1 \!+\!\! ~a_{1}\mathrm {Tr}\left( {\overline{\mathbf{H}_{i}\left ( q \right ) }\mathbf{F}} \right)} \right\rbrack \!-\! \mathrm{lnt_{u,k}} \!\!-\!\!1,
\end{eqnarray} 

In the same way, we set ${t = t}_{e}$, $x = 1 + ~a_{2}\mathrm{\mathrm{Tr}}\left( {\overline{\mathbf{H}_{j}\left ( q \right ) }\mathbf{F}} \right)$. Then, we can obtain
\begin{eqnarray}
	\begin{aligned}
		R_{e}\frac{\ln 2}{B} & = {\mathrm{ln}\left( {1 + ~a_{2}\mathrm {Tr}\left( {\overline{\mathbf{H}_{j}\left ( q \right ) }\mathbf{F}} \right)} \right)}\\
		&  = \underset{t_{e} > 0}{\mathrm{min}}~\varphi_{e}\left( \mathbf{F},t_{e} \right),
	\end{aligned}
\end{eqnarray}
where
\begin{eqnarray}
		{\varphi_{e}\left( \mathbf{F},t_{e} \right) \!\!=\!\! t}_{e}*\left\lbrack {1 \!\!+\!\! ~a_{2}\mathrm {Tr}\left( {\overline{\mathbf{H}_{j}\left ( q \right ) }\mathbf{F}} \right)} \right\rbrack - \mathrm{lnt_{e}} \!\!-\!\! 1,
\end{eqnarray} 

Based on Eqs.~(18) and (20), we can transform problem \textbf{P3} into problem \textbf{P4}, which is shown as follows
\begin{equation*}
	\begin{split}
		\textbf{P4:}&\mathop
		{\max\limits_{\mathbf{F},t_{u,k},t_{e}}\frac{{\min\limits_{k \in K}{\varphi_{u,k}\left( \mathbf{F},t_{u,k} \right)}} - \varphi_{e}\left( \mathbf{F},t_{e} \right)}{\frac{\ln 2}{B}(\mathrm{Tr}\left( \mathbf{F} \right) + P_\mathrm{BS} + P_\mathrm{IRS})}}\\
		\mathrm{s.t.}\quad &\begin{cases}
        C1':\mathbf{F} \in \mathcal{F},\\
		\hat{C}2:{\min\limits_{k \in K}{\varphi_{u,k}\left( \mathbf{F},t_{u,k} \right)}} - \varphi_{e}\left( \mathbf{F},t_{e} \right) \geq R_\mathrm{min}\frac{\ln 2}{B},\\
		C4:t_{u,k} \geq 0,t_{e} \geq 0.
     \end{cases}
	\end{split}
\end{equation*}
Based on Definition~1, the optimal solutions of $t_{u,k}$ and $t_{e}$, denoted by $t_{u,k}^{*}$ and $t_{e}^{*}$, can be derived as follows
\begin{eqnarray}
	\begin{cases}
		&t_{u,k}^{*} = \left\lbrack {1 + ~a_{1}\mathrm{Tr}\left\lbrack {\overline{\mathbf{H}_{i}\left ( q \right ) }\mathbf{F}} \right\rbrack} \right\rbrack^{- 1},\\
		
		&t_{e}^{*} = \left\lbrack {1 + ~a_{2}\mathrm{Tr}\left\lbrack {\overline{\mathbf{H}_{j}\left ( q \right ) }\mathbf{F}} \right\rbrack} \right\rbrack^{- 1}.
	\end{cases}
\end{eqnarray}
Since the $\textbf{P4}$ includes an embedded minimum operation, $\min\limits_{k \in K}$, making the problem intractable, we leverage SDR to transform this SEE maximazation problem into a relaxed form.  Plugging Eq.~(16) into the $\textbf{P4}$, and introducing a slack variable $\varepsilon_{\mathbf{F}} \leq \underset{k \in K}{\mathrm{min}}~\varphi_{u,k}$, we can transform problem $\textbf{P4}$ into problem $\textbf{P5}$ as follows
\begin{equation*}
	\begin{split}
		\textbf{P5:}&\mathop
		{\underset{\mathbf{F}}{~\mathrm{max}}~\varepsilon - \varphi_{e}\left( {\mathbf{F},t_{e}^{*}} \right) \! - \! \frac{\ln 2}{B}*\lambda\left( \xi \mathrm{Tr}\left( \mathbf{F} \right) \! + \! P_\mathrm{BS} \!+\! P_\mathrm{IRS} \right)}\\
		\mathrm{s.t.}\quad &\begin{cases}
        C1':\mathbf{F} \in \mathcal{F},\\
		\bar{C}2:\varepsilon_{\mathbf{F}}-\varphi_{e}\left( {\mathbf{F},t_{e}^{*}} \right) \geq R_\mathrm{min}\frac{\ln 2}{B},\\
		C5:\varphi_{u,k}\left( {\mathbf{F},t_{u,k}^{*}} \right) \geq \varepsilon_{\mathbf{F}}.
        \end{cases}
	\end{split}
\end{equation*}
The function $\varphi_{u,k}\left( {\mathbf{F},t_{u,k}^{*}} \right)$ and $\varphi_{e}\left( {\mathbf{F},t_{e}^{*}} \right)$ are respectively concave and convex, and every constraint exhibits convexity. Thus, we can resolve Problem $\textbf{P5}$ with standard convex optimization solvers.

Due to the constraints on $\mathbf{F}$, it is necessary to determine whether it satisfies the rank-1 condition and then conduct the corresponding solution: when $\mathrm{rank}\left( \mathbf{F} \right) = 1$, we perform the eigen decomposition $\mathbf{F}$ such that $\mathbf{F} = \mathbf{f}\mathbf{f}^{H}$ to acquire the vector $\mathbf{f}$; 
when $\mathrm{rank}\left( \mathbf{F} \right)\ne1$, we employ the Gaussian randomization to recover the vector $\mathbf{f}$.

\subsection{Optimizing the IRS phase shift $\overline{\mathcal{\mathbf{w}}}$  for Given $\mathbf{f}$ and ${q}$}

In the UAV-IRS assisted wireless communication system, we set $\mathbf{h}_{\mathbf{w},u_k}\left ( q \right ) = \mathbf{H}_{1}\left (q\right ) \mathbf{f}$, $\mathbf{h}_{\mathbf{w},e}\left (q\right )  = \mathbf{H}_{2}\left ( q \right ) \mathbf{f}$. Then, the SINRs of the users and Eve can be denoted as follows
\begin{eqnarray}
	\begin{aligned}
		\begin{cases}
			&\gamma_{\mathbf{w},u_k} = a_{1}\left| {{\overline{\mathcal{\mathbf{w}}}}^{H}\mathbf{h}_{\mathbf{w},u_k}\left ( q \right )} \right|^{2},\\
			
			&\gamma_{\mathbf{w},e} = a_{2}\left| {{\overline{\mathcal{\mathbf{w}}}}^{H}\mathbf{h}_{\mathbf{w},e}\left ( q \right )} \right|^{2}.
		\end{cases}
	\end{aligned}
\end{eqnarray}
In the same way, we denote $\mathbf{W} \!\!=\!\! \overline{\mathcal{\mathbf{w}}}\hspace{1pt}\overline{\mathcal{\mathbf{w}}}^{H}$, $\mathbf{H}_{\mathbf{w},u_k}\left ( q \right ) = \mathbf{h}_{\mathbf{w},u_k}\left ( q \right ){\mathbf{h}_{\mathbf{w},u_k}}\left ( q \right )^{H}$, and $\mathbf{H}_{\mathbf{w},e}\left ( q \right ) = \mathbf{h}_{\mathbf{w},e}\left ( q \right ){\mathbf{h}_{\mathbf{w},e}}\left ( q \right )^{H}$. Thus, the transformation from  problem $\textbf{P2}$ to problem $\textbf{P6}$ is presented as follows
\begin{equation*}
	\begin{split}
		\textbf{P6: \ }&{\max\limits_{\mathbf{W}}\left\{  \right.}{\underset{k \in K}{\min\{}{\frac{B}{\ln 2}{\mathit{\ln}{\left. \left( {1 + a_{1}\mathrm{Tr}\left( {\mathbf{H}_{\mathbf{w},u_k}\left ( q \right )\mathbf{W}} \right)} \right) \right\}\}}}}}\\&~~~~- \frac{B}{\ln 2}{\mathit{\mathrm{ln}}\left( {1 + ~a_{2}\mathrm{Tr}\left( {\mathbf{H}_{\mathbf{w},e}\left ( q \right )\mathbf{W}} \right)} \right)} - \lambda P_{tot}\\
		\mathrm{s.t.}\quad &\begin{cases}
        \check{C}2:\underset{k \in K}{\min\{}{\frac{B}{\ln 2}{\mathit{\ln}\left. \left( {1 + a_{1}\mathrm{Tr}\left( {\mathbf{H}_{\mathbf{w},u_k}\left ( q \right )\mathbf{W}} \right)} \right) \right\}}}\\~~~~~~~~~~- \frac{B}{\ln 2}{\mathit{\mathrm{ln}}\left( {1 + ~a_{2}\mathrm{Tr}\left( {\mathbf{H}_{\mathbf{w},e}\left ( q \right )\mathbf{W}} \right)} \right)} \geq R_\mathrm{min},\\
		\tilde{C}3:\mathbf{W} \succcurlyeq 0,\mathbf{W}_{nn} = 1,~n \in 1,2,\ldots,N,.\\
      \end{cases}
	\end{split}
\end{equation*}
Based on Definition 1, employing SDR method, we further introduce a slack variable satisfying $\varepsilon_{\mathbf{w}} \leq \underset{k \in K}{\mathrm {min}}\varphi_{\mathbf{w},u_k}$. Then, the problem $\textbf{P6}$ can be transformed as follows
\begin{equation*}
	\begin{split}
		\textbf{P7:}&{\max\limits_{\mathbf{W},t_{\mathbf{w},u_k},t_{\mathbf{w},e}}{\varepsilon_{\mathbf{w}} -}}\varphi_{\mathbf{w},e}\left( {\mathbf{W},t_{\mathbf{w},e}} \right) - \frac{\ln 2}{B}\lambda P_{tot}\\
		\mathrm{s.t.}\quad &\begin{cases}
        \breve{C}2:{\varepsilon_{\mathbf{w}} - \varphi}_{\mathbf{w},e}\left( {\mathbf{W},t_{\mathbf{w},e}} \right) \geq R_\mathrm{min}\frac{\ln 2}{B},\\
		\tilde{C}3:\mathbf{W} \succcurlyeq 0,\mathbf{W}_{nn} = 1,~n \in 1,2,\ldots,N,\\
        C6:\varphi_{\mathbf{w},u_k}\left( {\mathbf{W},t_{\mathbf{w},u_k}} \right) \geq \varepsilon_{\mathbf{w}},\\
      \end{cases}
	\end{split}
\end{equation*}
where 
\begin{eqnarray}
	\begin{aligned}
		\begin{cases}
			&{\varphi_{\mathbf{w},u_k}}\!\! = \!\!t_{\mathbf{w},u_k}*\left\lbrack {1 + ~a_{1}\mathrm{Tr}\left\lbrack {\mathbf{H}_{\mathbf{w},u_k}\mathbf{W}} \right\rbrack} \right\rbrack \!\!-\!\! \mathrm{ln}t_{\mathbf{w},u_k} \!\!-\!\! 1,\\
			
			&{\varphi_{\mathbf{w},e}} \!\!=\!\! t_{\mathbf{w},e}*\left\lbrack {1 + ~a_{2}\mathrm{Tr}\left\lbrack {\mathbf{H}_{\mathbf{w},e}\mathbf{W}} \right\rbrack} \right\rbrack \!\!-\!\! \mathrm{ln}t_{\mathbf{w},e} \!\!-\!\! 1,
		\end{cases}
	\end{aligned}
\end{eqnarray}

All objective and constraint expressions of problem $\textbf{P7}$ satisfy convexity, enabling efficient solution via convex optimization tools. Then, $t_{\mathbf{w},u_k}^*$ and $t_{\mathbf{w},e}^*$ can be derived as follows
\begin{eqnarray}
	\begin{aligned}
		\begin{cases}
			&t_{\mathbf{w},u_k}^{*} = \left\lbrack {1 \!+\! ~a_{1}\mathrm{Tr}\left\lbrack {\mathbf{H}_{\mathbf{w},u_k}\mathbf{W}} \right\rbrack} \right\rbrack^{- 1},\\
			
			&t_{\mathbf{w},e}^{*} = \left\lbrack {1 \!+\! ~a_{2}\mathrm{Tr}\left\lbrack {\mathbf{H}_{\mathbf{w},e}\mathbf{W}} \right\rbrack} \right\rbrack^{- 1}.
		\end{cases}
	\end{aligned}
\end{eqnarray}
Similarly, we calculate $\overline{\mathcal{\mathbf{w}}}$ either via eigenvalue decomposition or the Gaussian randomization of $\mathcal{\mathbf{W}}$, then use the definition of $\mathcal{\mathbf{\omega }}_{N}\mathord{=}e^{j\theta_{N}}$, 
$\mathcal{\mathbf{w}}^{H}\mathord{=}\left( {{\omega }_{1}, {\omega }_{2},\ldots,{\omega }_{N}} \right)$, and $ {\overline{\mathcal{\mathbf{w}}}}^{H} \!\!\!=\!\! e^{j{\mathfrak{w}}}\left (\mathbf{w}^{H},1\right) $ solve for corresponding reflection coefficients.

\subsection{Optimizing UAV Trajectory ${q}$ for Given $\mathbf{f}$ and $\overline{\mathcal{\mathbf{w}}}$}
We fix $\overline{\mathcal{\mathbf{w}}}$ and  $\mathbf{f}$ to optimize UAV Trajectory ${q}$. Thus, the SEE optimization problem $\textbf{P2}$ is reformulated by problem $\textbf{P8}$, shown as follows
\begin{equation*}
	\begin{split}
		\textbf{P8:}~&\mathop {\max\limits_{q}~}\min_{k \in K} \Bigg\{ 
		\frac{B}{\ln 2} \ln \left( 1 + a_{1} \left| {{\overline{\mathcal{\mathbf{w}}}}^{H}\mathbf{H}_{1}\left ( q \right ) \mathbf{f}} \right|^{2} \right)\Bigg\} \\
		& - \frac{B}{\ln 2} \ln \left( 1 + a_{2} \left|{{\overline{\mathcal{\mathbf{w}}}}^{H}\mathbf{H}_{2}\left ( q \right ) \mathbf{f}} \right|^{2} \right) 
		 - \lambda P_{tot}, \\
		\mathrm{s.t.}&\quad C2'
	\end{split}
\end{equation*}

We need to determine an optimal position for the UAV-mounted IRS within its available movement state space, amidst channel variations caused by high-speed UAV movement. To solve this issue, the trajectory optimization task for the UAV-assisted IRS is restructured into a discrete-time Markov Decision Process (MDP). Mathematically, this MDP is defined as a tuple $\mathcal{M}=( \mathcal{S}, \mathcal{A}, \mathcal{R})$, comprising the finite state space $\mathcal{S}$, the action space $\mathcal{A}$, and the reward space $\mathcal{R}$. The detailed definitions are presented as follows.

$1)$ State Space $\mathcal{S}$:~Since channels are generally time-varing, and reinforcement learning requires real-time observation of the environment state, the channel matrices $\mathbf{H}_{1}$, $\mathbf{H}_{2}$ and the UAV trajectory $q$ are incorporated into $\mathcal{S}$. The current environment state at time $t$, denoted by $s_t$, can be shown as follows
\begin{eqnarray}
	s_t = (\mathbf{H}_{1}(t),\mathbf{H}_{2}(t),{q(t)})\in \mathcal{S}.
\end{eqnarray}

$2)$ Action Space $\mathcal{A}$:~$\mathcal{A}$ refers to all feasible moving options for UAV displacement, i.e., $\mathcal{A}\!\!=\!\!\left \{(0,-1,0), (1,0,0), (0,1,0)\right \}$. We denote $a_t$ the location of UAV at the moment $t$, shown as follows
\begin{eqnarray}
	a_t = (x, y, z) \in \mathcal{A}.
\end{eqnarray}	
    
$3)$ Reward $\mathcal{R}$:~The aim of $\textbf{P8}$ is to maximize the SEE for the UAV-assisted IRS wireless communications system. We denote $r_t$ the reward function at the moment $t$,  shown as follows
\begin{eqnarray}
	r_t = \begin{cases}
\eta_{SE}(s_t, a_t) -\wp \|{q_{t+1}} - {q_t}\| ^2, & \text{if $C2'$ hold} \\
0, & \text{otherwise.}
\end{cases}
\end{eqnarray}
where $\eta_{SE}(s_t, a_t)$ is the SEE at the moment $t$. $\wp$ denotes the UAV movement penalty, and $q_{t+1}$ denotes the next UAV location.

To solve the formulated MDP for UAV-IRS trajectory optimization, we employ Deep Reinforcement Learning (DRL) algorithm \cite{AI-Enabled_Multi-QoS_Provisioning_for_6G_HRLLC}, i.e., DQN \cite{A_Novel_Movie_Recommendation_System_Based_on_Deep_Reinforcement_Learning_with_Prioritized_Experience_Replay}. In the DQN framework, the agent continuously interacts with the environment to learn an optimal action policy that maximizes the expected long-term cumulative reward \cite{Capacity_Maximization_in_RIS-UAV_Networks_A_DDQN-Based_Trajectory_and_Phase_Shift_Optimization_Approach}. Mathematically, after the current environmental state $s_t$ is observed, action $a'$ is selected by the agent based on strategy $\pi$. The corresponding action-action value function, denoted by $Q^*(s_t,a_t)$, can be defined as follows
\begin{eqnarray}
	Q^*(s_t,a_t)=\underset{\pi}{\mathrm{max}}~\mathbb{E}\!\left[R_t\Big|\,s_t,a_t\right],
\end{eqnarray}
where $R_t \!\!\!=\!\!\! \sum_{t'=0}^{\infty} \chi_{t'} r_{t+t'}$ is the discounted cumulative reward. Here, $r_{t+t'}$ indicates an immediate reward at $t+t'$, and$\chi_{t'}\!\in\![0,1)$ denotes the discount factor at $t'$ for reward delivery. By applying the Bellman optimality principle \cite{Performance_Evaluation_of_DQN_DDQN_and_Dueling_DQN_in_Heart_Disease_Prediction} to Eq.~(29), it can be reformulated as follows
\begin{eqnarray}
	Q^*(s_t,a_t)=\mathbb{E}\!\left[\chi\underset{a'\in\mathcal{A}}{\max}~Q(s',a')+r_t\,\Big|\,s_t,a_t\right],
\end{eqnarray}
where $s_{t+1}$ and $a_{t+1}$ denote the next state and action, respectively. The value $Q(s', a')\! =\! \mathbb{E}\left[ R_{t+1} \mid s_{t+1}=s', a_{t+1}=a', \pi \right]$ represents the network’s output when operating in state $s_{t+1}$. To approximate the optimal Q-function, a neural network with parameter $\vartheta$ is introduced, such that  $Q(s_t,a_t;\vartheta)\approx Q^*(s_t,a_t)$. 

\begin{figure}
	\centering
	\includegraphics[scale=0.35, trim=60bp 180bp 10bp 70bp, clip]{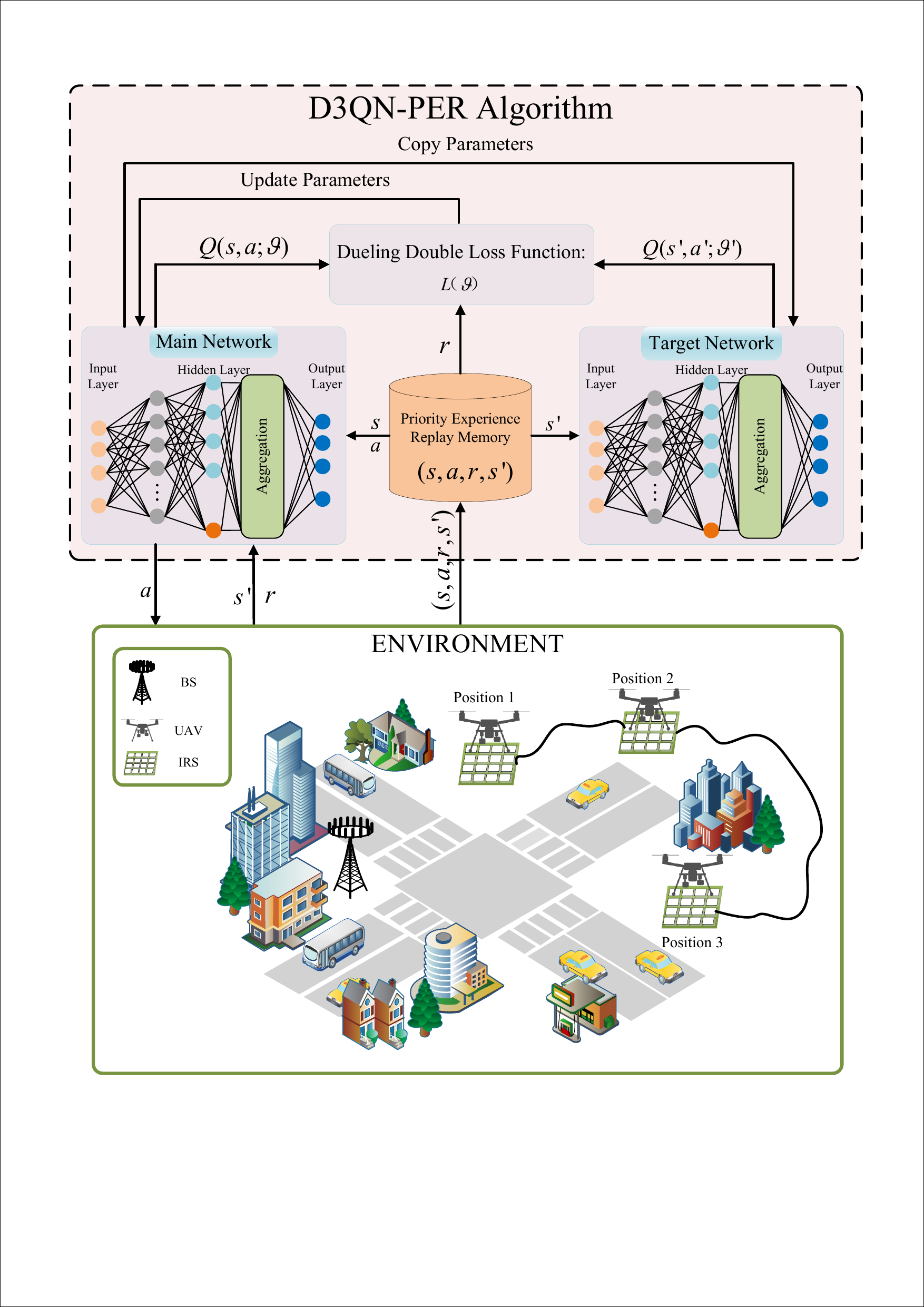}
	\caption{The general structure of D3QN-PER algorithm.}
	\vspace{-1pt}
\end{figure}

During training, each transition tuple $(s_t,a_t,r_t,s_t')$ generated  from the agent-environment interaction is stored in a replay buffer $\mathcal{D}$, from which mini-batches are randomly sampled to update the network. For a sampled transition $(s_t,a_t,r_t,s_t')$, the target value is defined as $y \!=\! r_t\!+\!\chi\underset{a'\in\mathcal{A}}{\max}Q(s_{t}',a';\vartheta')$. Accordingly, the expression for the loss function in the Q-network training process is as follows
\begin{eqnarray}
	\begin{aligned}
		&\!\!L(\vartheta) \!=\! \mathbb{E}_{\mathcal{D}} \left[ \left(y\!-\!Q(s_t, a_t;\vartheta) \right)^2 \right],
	\end{aligned}
\end{eqnarray}
where the online network parameter $\vartheta$ is updated by gradient descent, shown as follows
\begin{equation}
\vartheta \leftarrow \vartheta-\alpha \nabla_{\vartheta}L(\vartheta).
\end{equation}
Here, $\alpha$ denotes the learning rate. The target network's parameters, denoted by $\vartheta'$, are periodically updates via direct duplication of the online network, i.e.,
\begin{equation}
\vartheta' \leftarrow \vartheta.
\end{equation}

To evaluate the contributions of the state and action separately, we regard $V(s_t)$ as the state-value function and $A(s_t,a_t)$ as the advantage function, and decompose the Q-function accordingly. Thus, the dueling Q-value, denoted by $\tilde{Q}(s_t,a_t)$, can be derived as follows
\begin{equation}
\tilde{Q}(s_t,a_t)=V(s_t)+A(s_t,a_t)-\frac{1}{|\mathcal{A}|}\sum_{a'\in\mathcal{A}}A(s_t,a').
\end{equation}

To mitigate Q-value overestimation, we adopt the double DQN framework: the online network is used to select the action, while the target network is used to evaluate the selected action \cite{Capacity_Maximization_in_RIS-UAV_Networks_A_DDQN-Based_Trajectory_and_Phase_Shift_Optimization_Approach}. Thus, the target value, denoted by $\tilde{y}_t$, is rewritten as follows
\begin{equation}
\tilde{y}_t=r_t+\chi \tilde{Q}\!\left(s_t',\arg\max_{a'\in\mathcal{A}}\tilde{Q}(s_t',a';\vartheta);\vartheta'\right).
\end{equation}

\begin{algorithm}[!t]
	\caption{Integrated D3QN-PER-based Joint Alternating Optimization for SEE}
	\begin{algorithmic}[1]
		\REQUIRE Number of iterations $T$, The batchsize $B$, the discount factor $\chi$, the learning rate $\alpha$, the convergence accuracy $\delta$, minimum achievable secrecy rate ${R}_{\min }$, maximum transmission power of BS ${P}_{\max }$, the initial network parameters $\vartheta$ and $\vartheta'$, priority epsilon $\epsilon_{per}$.
		\ENSURE Optimal $Q^*$, $\mathbf{w}^*$, $\mathbf{f}^*$, $q^*$, $\lambda^*$
		\STATE Initialize networks $V(s;\vartheta_V)$ and  $A(s,a;\vartheta_A)$, with $\vartheta = [\vartheta_V, \vartheta_A]$. Initialize target dueling network $Q(\vartheta')$ with same structure
		\STATE Initialize prioritized replay buffer $\mathcal{D}$ and set exploration rate $\epsilon = 1$
		\STATE  Iitialization phase shifts of the IRS’s elements $\mathbf{w}(0)$, SEE $\mathbf{\lambda}(0)$, $i=1$		
		\WHILE{$\left\| \lambda (i)-\lambda (i-1) \right\|\ge \delta $}
        \STATE Initialize state $s$
		\STATE Solve \textbf{P5} with $\mathbf{w}^{*}(i-1)$, $q^{*}(i-1)$ to get $\mathbf{f}^{*}(i)$
		\STATE Solve \textbf{P7} with $\mathbf{f}^{*}(i)$, $q^{*}(i-1)$ to get $\mathbf{w}^{*}(i)$
        \STATE Solve \textbf{P8} with $\mathbf{f}^{*}(i)$, $\mathbf{w}^{*}(i)$ to get $q^{*}(i)$:
        \FOR{each step $t=1$ to $T$}
    	\STATE \hspace*{1em}Select action $a_t$ using $\epsilon$-greedy
		\STATE \hspace*{1em}Execute action $a_t$, reward $r_t$ and receive next state $s_{t+1}$
        \STATE \hspace*{1em}Calculate TD error $\delta_t$ and priority $p_t$ with Eq.~(36)\\
		\STATE \hspace*{1em}Store transition $(s_t,a_t,r_t,s_{t}',p_t)$ in replay memory $\mathcal{D}$\\
		\STATE \quad \text{\bf while} {Mini-batch $B$ is not full} \text{\bf do}
		\STATE \hspace*{2em}Sample transition $(s_j,a_j,r_j,s_j',p_j)$ from $\mathcal{D}$ with probability proportional to $p_j$
        \STATE \hspace*{2em}Add transition to Mini-batch $B$
	    \STATE \quad \text{\bf {end while}}
		\STATE \quad \text{\bf for} {each trainsition $(s_j,a_j,r_j,s_j')$ in Mini-batch $B$} \text{\bf do}
		\STATE \hspace*{2em}Calculate loss function $\tilde{L}_j(\vartheta)$ with Eq.~(37)
		\STATE \hspace*{2em}Calculate the gradient descent:
        \STATE \hspace*{3em}$\vartheta \leftarrow \vartheta - \alpha \nabla_\vartheta\tilde{L}_j(\vartheta)$
		\STATE \hspace*{2em}Update the target network:
        \STATE \hspace*{3em}$\vartheta' \leftarrow \vartheta$\\
	    \STATE \quad \text{\bf {end for}}
		\STATE Update $i=i+1$\\
		\ENDFOR		
        \ENDWHILE
	    \STATE \text{\bf Output:} optimal phase shifts of the IRS’s elements ${\mathbf{w}^{*}}$, optimal beamforming vector ${\mathbf{f}^{*}}$, optimal IRS Trajectory ${{q}^{*}}$, optimal SEE ${{\lambda}^{*}}$.
	\end{algorithmic}
\end{algorithm}

Furthermore, standard uniform sampling from $\mathcal{D}$ ignores the difference in transition informativeness and limits training efficiency. To mitigate this issue, we employ prioritized experience replay (PER), which prioritizes transitions with greater temporal-difference (TD) errors during sampling \cite{Enhanced_Routing_Scheduling_in_SDN_Using_Dueling_Double_DQN_with_Prioritized_Experience_Replay}. For a sampled transition $(s_t,a_t,r_t,s')$, the TD error, its corresponding transition priority, and sampling probability, denoted by $\delta_t$, $p_t$, and $P(t)$, are respectively given as follows
\begin{equation}
\left\{
\begin{aligned}
\delta_t &= \tilde{y}_t-\tilde{Q}(s_t,a_t;\vartheta), \\
p_t &= \left(|\delta_t|+\epsilon_{\mathrm{per}}\right)^{\mathfrak{d} }, \\
P(t) &= \dfrac{p_t}{\sum_{k\in\mathcal{D}}p_k},
\end{aligned}
\right.
\end{equation}
where $\epsilon_{per}$ refers to a small positive constant for avoiding zero priority, and $\mathfrak{d} $ regulates the prioritization level. The final loss function, denoted by $\tilde{L}(\vartheta)$, is rewritten as follows
\begin{equation}
\tilde{L}(\vartheta)=\mathbb{E}_{(s_t,a_t,r_t,s_t')\sim P(\mathcal{D})}
\left[\ell_t\left(\tilde{y}_t-\tilde{Q}(s_t,a_t;\vartheta)\right)^2\right].
\end{equation}
where $\ell_t$ is the sampling weight designed to correct the deviation caused prioritized sampling. The proposed D3QN-PER algorithm facilitates robust and highly efficient learning in high-mobility UAV scenarios.

Algorithm $1$ provide the SEE maximization algorithm based on D3QN-PER and Alternating Optimization (AO) method. The algorithm's complexity is dominated by the AO method, including SDR-based beamforming ($\mathcal{O}(I_1 M^{3.5})$), IRS phase shift ($\mathcal{O}(I_2 N^{3.5})$) optimizations, and the D3QN-PER-based UAV trajectory optimization ($\mathcal{O}(TBD)$). The overall complexity is $
\mathcal{O}\left(I_{\rm AO}\left(I_{1} M^{3.5} + I_{2} N^{3.5} + TBD\right)\right)$,
where $I_{\rm AO}$ is the number of AO iterations, $I_1, I_2$ are the convex solver iterations, and $T, B$, and $D$ are the training episodes, mini-batch size, and neural network parameter dimension, respectively.

\section{Performance Evaluation}
We use MATLAB-based simulations to validate and evaluate our proposed D3QN-PER-based joint optimization algorithm for UAV-IRS-assisted LAE wireless networks. First, the SEE is plotted versus the maximum transmit power and minimum transmit rate in Figs.~3 and 4, respectively. Next, Fig.~5 evaluates the SEE for different benchmarks algorithm. Then, Fig. 6 shows the impact of UAV positions on SEE. To analyze the UAV trajectory solution based on the D3QN-PER algorithm, the UAV search paths with different initial points are shown in Figs.~7 and 8. 
Furthermore, the proposed D3QN-PER-based algorithm's performance is verified by camparing the Dueling DQN, DQN, DDQN, DDQN-PER, and D3QN-PER in Figs.~9 and 10. The simulation parameters are set as: the number of reflective elements on the IRS $N=100$, and the transmitting antennas equipped on the BS $M=8$.

We assume that the BS, Eve and two legitimate users are on the same horizontal line, where their coordinates are $(0,0,0)$, $(60,40,0)$, $(90,0,0)$ and $(100,0,0)$, respectively. The movable IRS is kept at the same vertical height, where the coordinates of the IRS are $(x_{a} ,y_{a},z_{a} )$. For the indirect links, i.e., BS$\to$IRS and IRS$\to$users, the path-loss is model as $35.6+22.0\lg d$, whereas the direct BS$\to$users link follows $32.6+36.7\lg d$, with $d$ representing the separation distance of connected devices.

\indent Figure 3 shows SEE maxization under different maximum BS transmit power values. As observed from the simulation results shown in Fig.~3, the SEE of the UAV-assisted IRS securing wireless communications system exhibits a decreasing trend when the BS maximum transmission power $P_{\mathrm {max}}$ increases. This phenomenon can be attributed to two key factors. On one hand, with an increase in $P_{\mathrm {max}}$, the energy consumed by auxiliary circuits, i.e., signal processing modules, power amplifiers gradually rises, which directly contributes to the degradation of SEE. On the other hand, beyond a specific power threshold, the growth rate of energy consumption surpasses the improvement in transmission performance. The SEE is positively correlated with the number of IRS reflecting units grows. This trend is attributed to the enhanced receive gain of the BS array achieved by employing more reflecting elements, which directly improves the system's SEE. Notably, the SEE achieved without IRS assistance is significantly lower. These results demonstrate that integrating an IRS can effectively enhance the SEE. Furthermore, the proposed scheme is superior to benchmark “initial trajectory”.
It demonstrates that optimizing the UAV position can result from huge performance gains.

\begin{figure}
    \centering
    \setcounter{tempfig}{\value{figure}} 
    \includegraphics[scale=0.5]{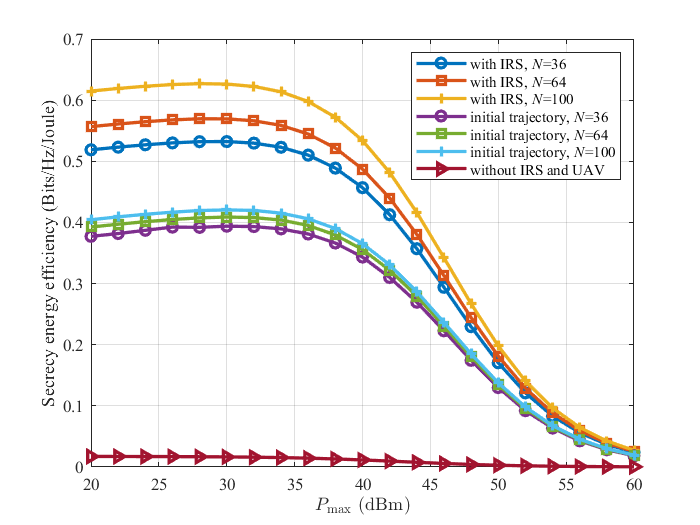}
    \caption{SEE versus $P_\mathrm{max}$.}
    \vspace{-1pt}
\end{figure}
\begin{figure}
    \centering
    \stepcounter{figure} 
    \includegraphics[scale=0.5]{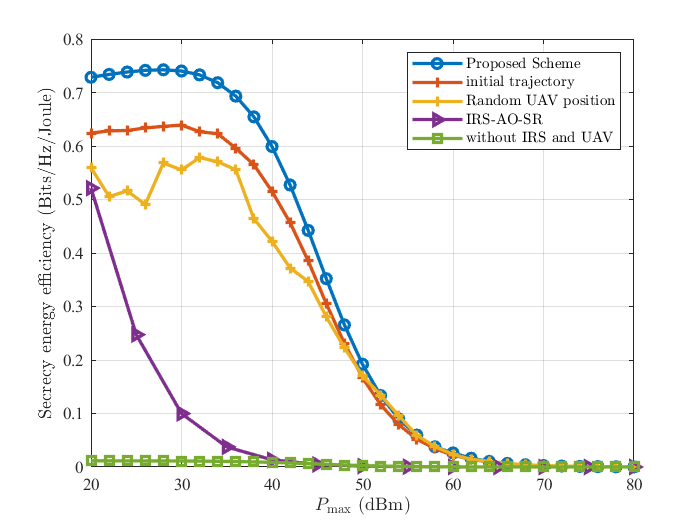}
    \caption{SEE versus $P_\mathrm{max}$ for different benchmarks.}
    \vspace{-1pt}
\end{figure}
\begin{figure}
    \centering
    \setcounter{figure}{\value{tempfig}} 
    \stepcounter{figure} 
    \includegraphics[scale=0.5]{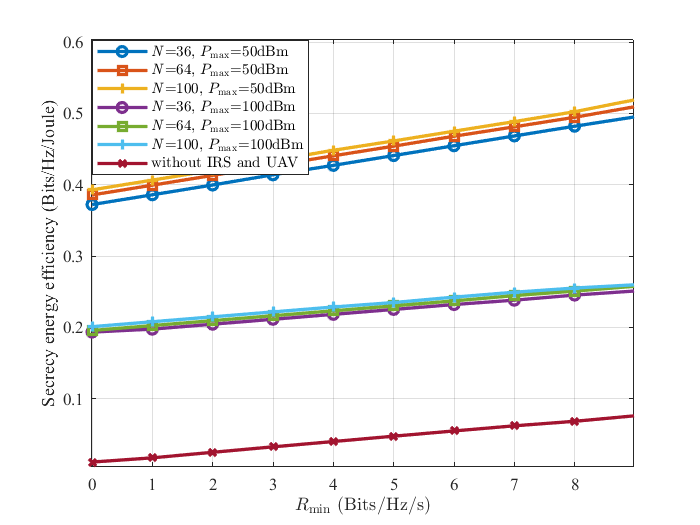}
    \caption{SEE versus $R_\mathrm{min}$.}
    \vspace{-1pt}
\end{figure}
\begin{figure}
    \centering
    \stepcounter{figure} 
    \includegraphics[scale=0.5]{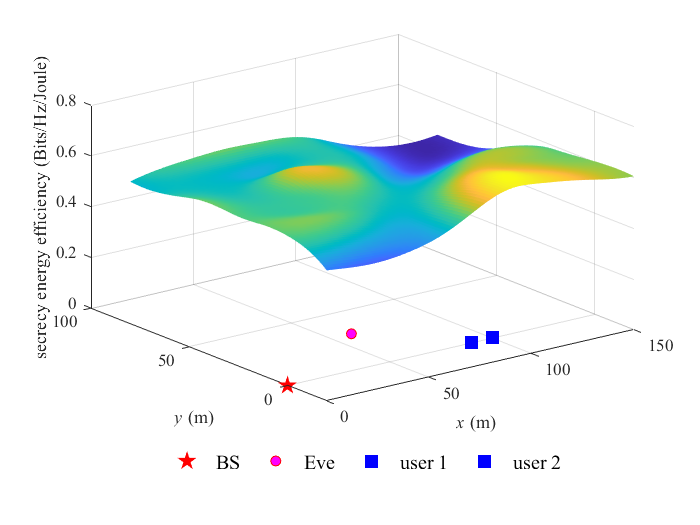}
    \caption{Impact of UAV trajectory on SEE.}
    \vspace{-1pt}
\end{figure}

\indent Figure 4 depicts SEE as the minimum achievable secrecy rate $R_{\mathrm {min}}$ varies in UAV-IRS-assisted LAE wireless network. As illustrated in Fig.~4, the SEE increases with the $R_{\mathrm {min}}$. This trend arises since achieving a stricter minimum secrecy rate necessitates a larger rate gap between the legitimate and the eavesdropping links, which raises the suppression requirement for eavesdropping link. Meanwhile, with the help of the IRS and UAV, the legitimate channel can be improved and the eavesdropping channel can be weakened with only a small extra energy cost. Accordingly, the secrecy rate increases faster than the power consumption, and thus the SEE increases as the minimum secrecy rate increases. By contrast, systems without IRS and UAV exhibit a significantly reduced secrecy energy efficiency. This result further demonstrates the considerable performance improvement afforded by IRS integration. The SEE of the UAV-IRS assisted wireless communications network grows with a larger number of reflecting elements.
Moreover, the system generally achieves better SEE under $P_{\mathrm {max}}=50\mathrm {dBm}$ than under $P_{\mathrm {max}}=100\mathrm {dBm}$, which aligns with the earlier observation that excessive maximum transmission power leads to efficiency degradation.

\indent Figure 5 compares our proposed scheme with four benchmark schemes. The“Random UAV Position” baseline optimize the transmit beamforming and IRS phase shift while the UAV opsition is random. The “without IRS” benchmark excludes the IRS from the system, thereby only jointly designing the UAV trajectory and transmit beamforming. The “initial trajectory” benchmark denotes the UAV is fixed at its original flight position. The benchmark “IRS-AO-SR” represents the maximum secrecy rate (SR) using the AO algorithm in the IRS-assisted wireless communication network. Fig.~5 illustrates the SEE variation against $P_{\mathrm {max}}$. The results show that the SEE exhibits a trend of decreasing as the $P_{\mathrm {max}}$ increases, eventually converging toward zero. This is because within a certain range of maximum transmission power, increasing the transmit power improves the SEE. By enhancing transmission power the received signal strength, which reduces transmission errors and retransmissions, thus enabling more effective information transmission per unit energy consumed. However, when the maximum transmission power exceeds this range and continues to increase, the network SEE gradually decreases. This decrease is primarily attributed to the increased additional energy losses in circuit components, i.e., the efficiency degradation of power amplifiers, and the adverse effects of nonlinear distortion caused by excessive power, which collectively offset the benefits of enhanced signal strength. For the UAV-IRS-assisted LAE wireless communications networks, relatively high transmission power may also lead to an increased risk of signal leakage and a rise in the possibility of eavesdropping, thus reducing the security of the system. 
\begin{figure}
    \centering
    \setcounter{tempfig}{\value{figure}} 
  \includegraphics[scale=0.58, trim=10bp 70bp 10bp 60bp, clip]{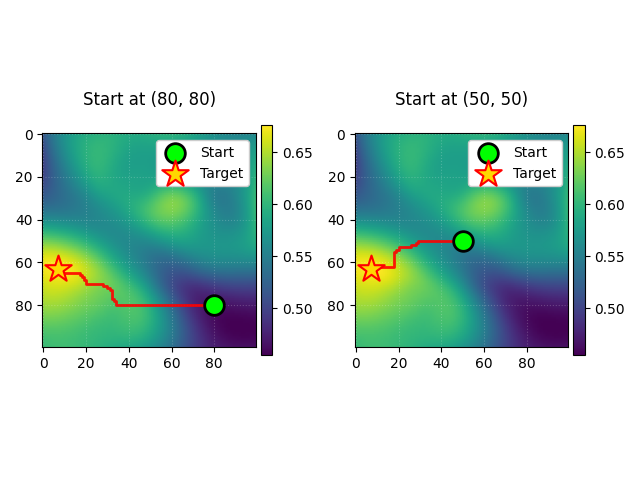}
    \caption{UAV trajectory searching paths under different start points, $P_{\mathrm {max}}= 35$.}
    \vspace{-1pt}
\end{figure}
\begin{figure}
    \centering
    \stepcounter{figure} 
   \includegraphics[scale=0.65, trim=10bp 5bp 10bp 30bp, clip]{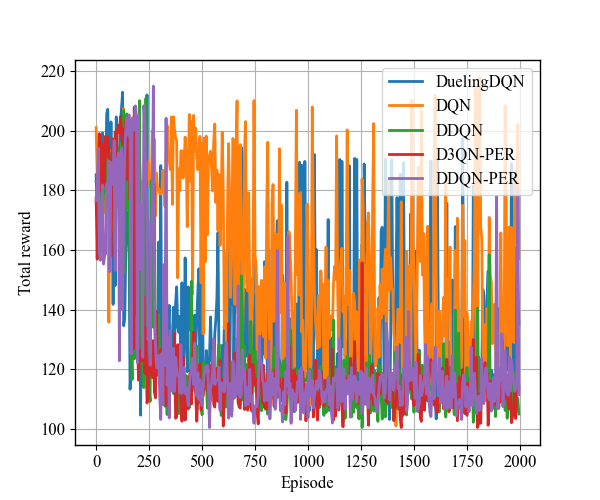}
    \caption{Comparisions of different DRL on reward.}\label{fig:right1}
    \vspace{-1pt}
\end{figure}
\begin{figure}
    \centering
    \setcounter{figure}{\value{tempfig}} 
    \stepcounter{figure} 
 \includegraphics[scale=0.58, trim=10bp 70bp 10bp 60bp, clip]{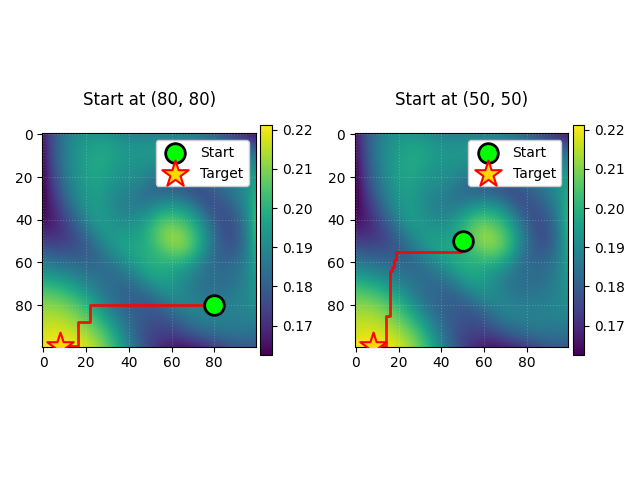}
    \caption{UAV trajectory searching paths under different start points, $P_{\mathrm {max}}= 50$.}
    \vspace{-1pt}
\end{figure}
\begin{figure}
    \centering
    \stepcounter{figure} 
   \includegraphics[scale=0.65, trim=10bp 5bp 10bp 30bp, clip]{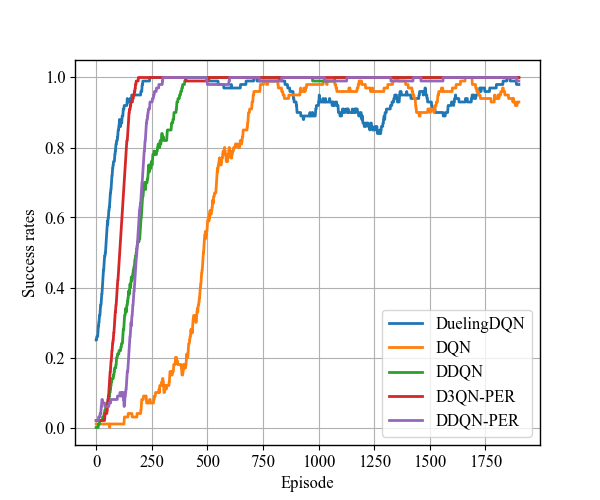}
    \caption{Comparisions of different DRL on success rate.}\label{fig:right2}
    \vspace{-1pt}
\end{figure}
Specifically, the SEE achieved by our proposed scheme is considerably higher than those of “Random UAV Position”, “initial trajectory” and “Without IRS” benchmarks, although the performance gap narrows at higher transmit power levels. It can be concluded that jointly optimizing the transmit beamforming, the IRS phase shift, and the UAV trajectory is crucial for improving secure transmission performance. For the “IRS-AO-SR” scheme, it remains almost unchanged and close to zero. This illustrates that the proposed optimization scheme can suppress eavesdropping effectively compared to the sum transmission rate maximization scheme.

\indent Figure 6 illustrates the variation of the system's SEE as the IRS-equipped UAV adjusts its trajectory within a planar airspace. Observing from Fig.~6, the SEE achieves its optimal value with the IRS positioned adjacent to the legitimate user. This is because proximity to legitimate users boosts their received signal strength, while decreasing leakage to the eavesdropper. For a given level of system energy consumption, the IRS achieves the higher secrecy rate in this region due to the favorable SINR for legitimate communications. In addition, SEE degrades correspondingly as the UAV-aided IRS trajectory extends further from the BS. 
This phenomenon can be attributed to the degraded signal power received by the IRS, which stems from the increased path loss over the BS$\to$IRS link, subsequently degrading the SEE for the UAV-IRS assisted LAE wireless communications network. Furthermore, Fig.~6 reveals a local maximum point of SEE in the vicinity of the eavesdropper. Actually, the local maximum occurs in the midpoint region of the XOZ plane between BS and users. At this specific location, the IRS is locally optimally positioned to balance signal reception from the BS and signal reflection to the legitimate user. On the other hand, although the IRS at this region maintains relatively high received power from the nearby BS, the signal leakage to the co-located eavesdropper is offset by the gains from the balanced relay position. Thus, the combined effect of efficient signal reception and controlled interference results in the formation of this local maximum.

\indent Figure 7 shows the training process of the D3QN-PER algorithm for optimizing the IRS trajectory under the constraint of maximum transmission power $P_{\mathrm {max}}=35\mathrm {dBm}$. Simulation results demonstrate that the model trained using D3QN-PER consistently converge to the position of maximum SEE regardless of the UAV's initial position. Similarly, Fig.~8 presents the trajectory optimization results under $P_{\mathrm {max}}=50\mathrm {dBm}$. Under this higher power constraint, the trained model retains the ability to guide the UAV to the global maximum SEE position from any initial location. Notably, even if the UAV temporarily approaches a local optimal region during its flight, the algorithm can still drive the trajectory toward the global maximum SEE, indicating a strong capability to escaping local optima. A comparative analysis between the two figures reveals that the system generally achieves better SEE under $P_{\mathrm {max}}=35\mathrm {dBm}$ than under $P_{\mathrm {max}}=50\mathrm {dBm}$, which aligns with the earlier observation that excessive maximum transmission power leads to SEE degradation.

\indent Figures 9 and 10 depict the training processes of deep reinforcement learning algorithms, i.e., Dueling DQN, DQN, DDQN, D3QN-PER, and DDQN-PER, in the UAV-IRS-assisted LAE wireless network, focusing on the evolution of reward values and success rates during training. As demonstrated in Fig.~9, D3QN-PER shows a fluctuating upward trend in early training. Although fluctuations in the later stage, it relatively stabilizes within a high reward interval. This indicates that as training episodes, D3QN-PER effectively learns and optimizes strategies, and continuously obtain stable and substantial rewards. In contrast, Dueling DQN, DQN, DDQN, and DDQN-PER fluctuate significantly, especially in early-to-middle training. In the later stage, these algorithms fail to converge to a relatively stable, high reward level like D3QN-PER. Compared with other algorithms, D3QN-PER has better stability in reward acquisition and can more stably maintain a higher reward interval in the later stage. This means that in strategy optimization and environmental adaptation, D3QN-PER can more effectively learn high-quality strategies, enabling the agent to obtain returns more reliably and efficiently in task execution, with better algorithm robustness and learning effect. Observing Fig.~10, D3QN-PER exhibits a rapid increase in average value at the initial training stage. After approximately 250 episodes, it stabilizes at a high level and maintains this performance despit minor subsequent fluctuations. This behavior indicates that D3QN-PER can quickly learn high-quality strategies, achieving a highly stable state of average value. Dueling DQN rises rapidly in the early stage, but its stabilized average value is lower than that of D3QN-PER. DQN increases slowly initially, with its average value remaining lower than the both Dueling DQN and D3QN-PER throughout the process, and stabilizing at a low level in the later stage. The early rising speed of DDQN is slower than that of Dueling DQN and D3QN-PER, and its stable level in the later stage is inferior to D3QN-PER. Similarly, DDQN-PER rises slowly initially, and although it improves and stabilizes in the later stage, its performance remains lower than that of D3QN-PER. In terms of learning speed, D3QN-PER rapidly enhances the average value in the initial stage, showing higher efficiency in learning the environment and strategies. It can explore valuable strategies more quickly and achieve high performance within a shorter training time. Regarding stability, D3QN-PER maintains the average reward at a high level, whereas other algorithms converge at lower levels or with longer time costs. These results indicate that D3QN-PER has better stability and convergence, enabling  more reliably support reinforcement learning tasks and improve the agent performance.
\section{Conclusion}\label{sec:con}
This paper considered a LAE secrecy wireless communications system. We maximized the SEE over the UAV-IRS assisted communications system with an eavesdropper under a perfect Channel State Information (CSI) model. Dinkelbach's method was used to solve the nonconvex fractional existing in SEE. By introducing relaxation variables and auxiliary variables, the two objective functions were changed from the original non-convex functions to convex approximate forms. For the UAV trajectory optimization subproblem, a D3QN-PER algorithm was proposed to address the slow and unstable convergence issues of conventional DQN. A joint AO algorithm was employed to obtain a feasible solution for our SEE function. The simulation results indicated that our designed scheme can realize secure communications while sustaining low power consumption for the UAV-IRS assisted wireless communications system. Comparative analyses further demonstrated that the D3QN-PER-based algorithm outperforms state-of-the-art learning methods, confirming its superiority in improving SEE for UAV-IRS-assisted LAE wireless communications networks.

\bibliographystyle{IEEEtran}
\small
\bibliography{UAV_IRS.bib}
\end{document}